\documentclass[11pt]{article}
\usepackage{geometry}                % See geometry.pdf to learn the layout options. There are lots.
\pdfoutput=1

\usepackage{setspace}
\usepackage{graphicx}
\usepackage{amssymb}
\usepackage{bm}
\usepackage{epstopdf}
\usepackage{dblaccnt}
\newcommand{\ra}[1]{\renewcommand{\arraystretch}{#1}}

\usepackage[square,authoryear]{natbib}
\usepackage[titletoc,title]{appendix}
\usepackage[font=small,labelfont=bf]{caption}
\usepackage{placeins}
\usepackage{booktabs}

\title{Isostasy with Love: II Airy compensation arising from viscoelastic relaxation}
\author{Mikael Beuthe\\
\it Royal Observatory of Belgium, Brussels, Belgium\\
mikael.beuthe@observatory.be
 }    
\date{}                                             % Activate to display a given date or no date

\begin{document}

\maketitle

\begin{abstract}

In modern geodynamics, isostasy can be viewed either as the static equilibrium of the crust that minimizes deviatoric stresses, or as a dynamic process resulting from the viscous relaxation of the non-hydrostatic crustal shape.
Paper~I gave a general formulation of Airy isostasy as an elastic loading problem solved with Love numbers, and applied it to the case of minimum stress isostasy.
In this sequel, the same framework is used to study Airy isostasy as the long-time evolution of a viscoelastic shell submitted to surface and internal loads.
Isostatic ratios are defined in terms of time-dependent deviatoric Love numbers.
Dynamic isostasy depends on the loading history, two examples of which are the constant load applied on the surface in the far past and the constant shape maintained by addition or removal of material at the compensation depth.
The former model results in a shape decreasing exponentially with time and has no elastic analog, whereas the latter (stationary) model is equivalent to a form of elastic isostasy.
Viscoelastic and viscous approaches are completely equivalent.
If both load and shape vary slowly with time, isostatic ratios look like those of the stationary model.
Isostatic models thus belong to two independent groups: the elastic/stationary approaches and the time-dependent approaches.
If the shell is homogeneous, all models predict a similar compensation of large-scale gravity perturbations.
If the shell rheology depends on depth, stationary models predict more compensation at long wavelengths, whereas time-dependent models result in negligible compensation.
Mathematica and Fortran codes are available for computing the isostatic ratios of an incompressible body with three homogeneous layers.

\end{abstract}

\vspace{\stretch{1}}
\newpage

{\small
\tableofcontents
\newpage
\listoffigures
\listoftables
}

\newpage

%%%%%%%%%%%%%%%%%%%%%%%%%%%%%%%%%%%%%%%%%%%%%%%%%%%%%%%%%%%%%%%%%%%%%%%%%%%%%%%%%%%%%%%%%%%%%%%%%%%%%%%
%%%%%%%%%%%%%%%%%%%%%%%%%%%%%%%%%%%%%%%%%%%%%%%%%%%%%%%%%%%%%%%%%%%%%%%%%%%%%%%%%%%%%%%%%%%%%%%%%%%%%%%
%%%%%%%%%%%%%%%%%%%%%%%%%%%%%%%%%%%%%%%%%%%%%%%%%%%%%%%%%%%%%%%%%%%%%%%%%%%%%%%%%%%%%%%%%%%%%%%%%%%%%%%
\section{Introduction}

At large scale, most topography on Earth (bathymetry included) is in isostatic equilibrium, meaning that surface loads are buoyantly supported by subsurface mass anomalies, due to crustal thickness variations or density variations within the crust and lithosphere, in such a way that stresses are hydrostatic below a constant depth called the compensation depth \citep{phillips1980}.
Textbook examples are the average depth of the oceans, the elevation of the Himalayas, and the subsidence of ocean floor away from midocean ridges, respectively explained by the thin oceanic crust (compared to thick continental crust), crustal thickening under mountain belts, and density increase due to plate cooling \citep{turcotte2014,fowler2005}.
In general, the isostatic concept is applied very simply, by partitioning the crust into vertical columns floating independently in a fluid and in mechanical equilibrium.
This picture is known to be oversimplified both at short and long wavelengths \citep{lambert1930,jeffreys1932,jeffreys1943,vening1946,dahlen1982}.
While the short-scale issue can be handled by combining isostasy with lithospheric flexure (that is flexural isostasy), the large-scale inadequacy does not appear clearly in Earth data because mantle flow becomes the dominant factor in large-scale geoid anomalies \citep{hager1985a,hager1985b}.
Moreover the contribution of mantle flow to the long-wavelength shape (called dynamic topography) is comparable to isostatic topography \citep{hoggard2016,davies2019,flament2019}.

On terrestrial-type planetary bodies, isostatic equilibrium occurs locally but has not yet been found to operate globally: geologic provinces with highly compensated gravity have been identified on Venus \citep{smrekar1991,simons1997}, Mars \citep{mcgovern2002,mcgovern2002errata}, and the Moon \citep{sori2018a}.
Nevertheless, predictions of dynamic topography on Venus depend, as on Earth, on the computation of isostatic topography at long wavelengths (e.g.\ \citet{pauer2006}).
Isostasy found a new domain of application in icy satellites with subsurface oceans, where the ice-to-water transition embodies the isostatic requirement of a hydrostatic layer below the depth of compensation.
Global isostasy thus seems likely in that setting; gravity data suggest that it operates on Enceladus \citep{iess2014,mckinnon2015,beuthe2016,hemingway2018}, Dione \citep{beuthe2016,zannoni2020}, and Titan \citep{durante2019}, whereas the case of Europa is not yet settled for lack of data \citep{nimmo2010}.
On the dwarf planet Ceres, gravity and shape data suggest nearly global isostatic compensation, although the negative gravity-shape correlation at degree and order two is incompatible with global isostasy \citep{ermakov2017,park2020}.
Icy moons and dwarf planets thus raise again the question of defining isostasy at all scales.

Attempts at improving classical isostasy belong to three categories.
The first one includes arguments that some models of classical isostasy are better than others.
For example, \citet{hemingway2017} argue that classical isostasy assuming equal lithostatic pressure at the compensation depth should be preferred to the most common model of equal mass columns, whereas \citet{cadek2019} claim that isostasy based on a principle of equal weight is a better approximation.
Since neither `equal pressure' nor `equal weight' isostasy can be derived from fundamental physical principles, the error of each model must be estimated by building a more complete model of isostasy (see \citet{beuthe2021}, hereafter called Paper~I).
In the second category,  isostasy is treated as a problem of elastic equilibrium, which is determined so that deviatoric stresses are minimized within the crust or lithosphere \citep{jeffreys1959,dahlen1982,beuthe2016}.
The concept of `minimum stress isostasy' originates in the observation that isostasy is a natural state in the evolution of a mechanical system in response to an applied stress field, because a mechanical system will always tend to respond in such a manner as to minimize stress \citep{phillips1980}.
Paper~I implemented this idea by formulating isostasy as the response, fully described by Love numbers, of a thick elastic shell floating on a fluid layer to the combined action of surface and internal loads.
In the third category, isostasy naturally arises from viscous relaxation.

\begin{figure}
   \centering
    \includegraphics[width=\textwidth]{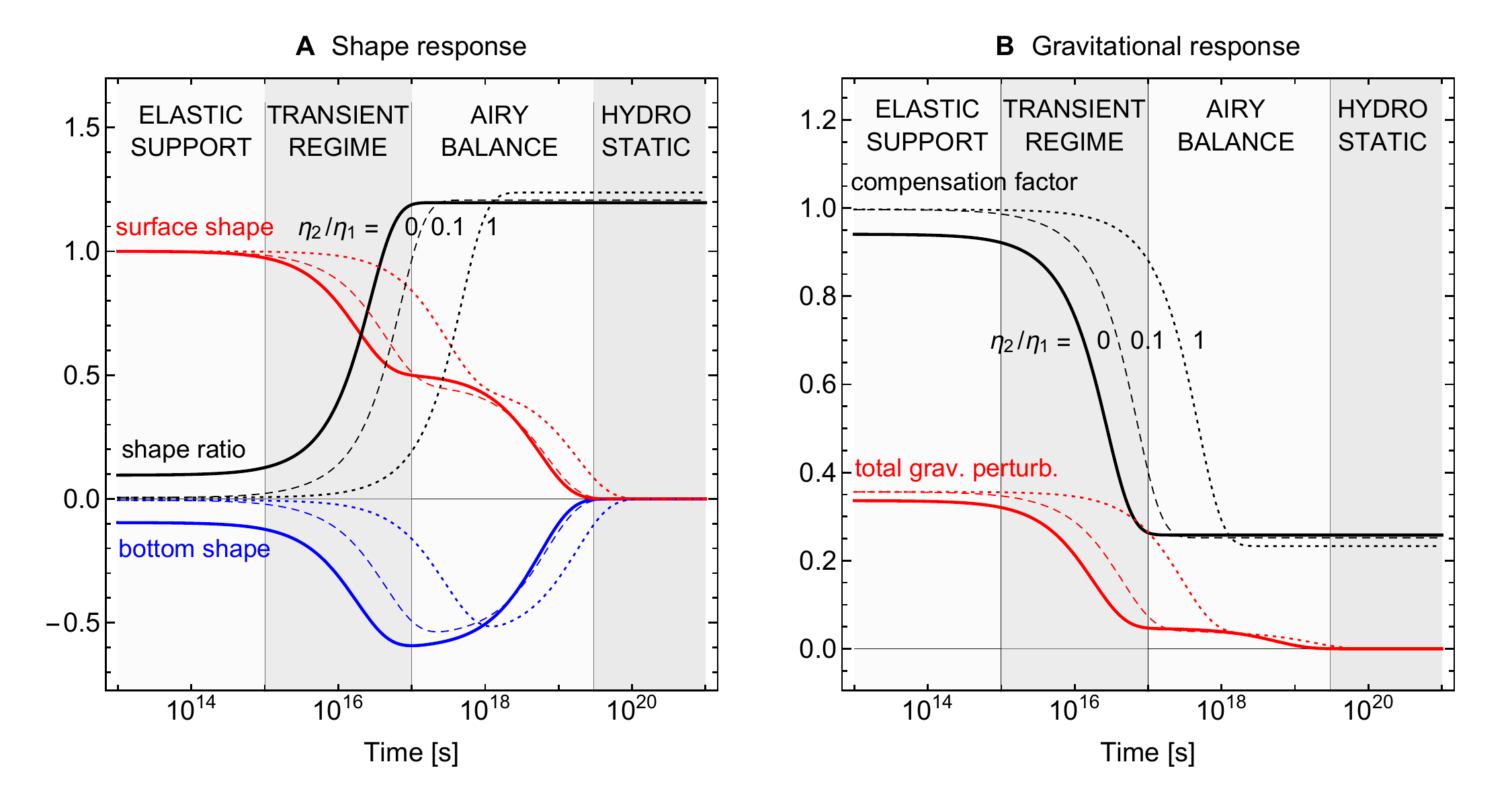}
   \caption[Viscoelastic relaxation of a two-layer spherical body]{Viscoelastic relaxation of a two-layer spherical body: (A) shape of interfaces and shape ratio; (B) perturbation of gravitational potential and compensation factor.
   A degree-2 load is applied at the surface at $t=0$ and remains constant.
    The shapes and the gravitational perturbation are normalized so that the initial surface shape is unity.
   The shape ratio (shown in absolute value) and compensation factor are defined in Section~\ref{SectionLoading}.
     The viscosities of the bottom and top layers are in the ratio $\eta_2/\eta_1=0$ (solid), 0.1 (dashed), or 1 (dotted).
    Shaded areas distinguish the four successive stages for the case of a fluid bottom layer ($\eta_2/\eta_1=0$): elastic support, transient regime, Airy balance, and hydrostatic equilibrium.
    This example is based on the two-layer Ceres model specified by $\rho_1=1286\rm\,kg/m^3$; $\rho_2=2434\rm\,kg/m^3$; $R_1=469.7\rm\,km$; $R_2=428.8\rm\,km$; $\mu_1=3.5\rm\,GPa$; $\eta_1=10^{25}\rm\,Pa.s$ \citep{ermakov2017}.
   }
   \label{FigShapeGravTime}
\end{figure}

A common feature of classical isostasy and minimum stress isostasy is that they avoid the question of the loading history.
This feature is simultaneously an advantage, because the loading history is most often unknown, and a drawback, because it is not clear whether plausible physical processes can lead to that outcome, and whether the isostatic balance is stable.
Faulting of a brittle lithosphere can contribute to isostasy, but it is more likely that isostatic deformations at depth are due to creep.
It has long been known that a perturbed viscous layer above a less viscous layer first relaxes to a state looking like Airy isostatic balance (at least for long-wavelength perturbations), before decaying slowly to hydrostatic equilibrium \citep{ramberg1968,phillips1980,solomon1982,zhong1997}.
Figure~\ref{FigShapeGravTime} shows an example of the relaxation of a degree-two perturbation on the dwarf planet Ceres.
The body is modelled as a two-layer viscoelastic sphere, with a more viscous shell of density $\rho_1$ surrounding a less viscous mantle/core of density $\rho_2$.
The initial stage of elastic support is followed by a transition to Airy balance, in which the shape ratio (i.e.\ the ratio of the shape of the bottom of the shell to the surface shape) becomes constant and approximately equal to $-(\rho_2-\rho_1)/\rho_1$.
The system then slowly evolves toward hydrostatic equilibrium, unless the evolution is stopped at the intermediate isostatic stage because stresses have fallen below the creep threshold \citep{phillips1980}.

In this context, Airy isostasy is a particular stage of the more general process of isostatic adjustment, which describes the time-dependent response of a planetary body to the modification of surface loads.
On Earth, the best-known example of isostatic adjustment is the postglacial rebound of Scandinavia \citep{steffen2011}, with similar rebounds occurring in Canada and Antarctica.
These observed changes in surface elevation are now seen as one manifestation of Glacial Isostatic Adjustment (GIA) among others, such as global sea-level change, gravity field variations, fluctuations of the Earth's rotation axis and length of day etc.\ (see reviews by \citet{mitrovica2011} and \citet{whitehouse2018}; theoretical aspects are covered in \citet{sabadini2016}).
Given the abundance of time-dependent data (sea level, GPS, and gravity), it has become possible to build global models of glacial loading \citep[e.g.][]{peltier2004,lambeck2014,peltier2015}, constraining at the same time the global sea level history \citep[e.g.][]{kopp2009} and the mantle viscosity \citep[e.g.][]{mitrovica1996,lau2016}.
Interior models have become complex, with viscosity varying not only with depth but also laterally \citep[e.g.][]{a2013}.
GIA data, however, are not sufficient to invert for lateral variations of viscosity, and a priori knowledge about them is lacking \citep{whitehouse2018}.

Beyond Earth, isostatic adjustment of impact craters and basins (and sometimes the global shape) is widely used to constrain the rheology of planets and moons.
Following early works \citep{parmentier1981,solomon1982,passey1983,grimm1988,thomas1988b}, topographic relaxation is now systematically studied on all large bodies of the solar system:
the Moon \citep{zhong2000,mohit2006,kamata2012,kamata2013,qin2018}, Mars \citep{zhong2002,pathare2005,mohit2007}, Mercury \citep{mohit2009}, dwarf planets \citep{fu2014,bland2016,ermakov2017,fu2017,sori2017},
Galilean satellites \citep{dombard2000,dombard2006,bland2017}, and Saturnian satellites \citep{robuchon2011,bland2012,white2013,white2017}.

In the theory of isostatic adjustment, isostasy refers to the initial state (before loading or unloading) and to the final stationary state reached after an infinite time.
The infinite time response of the system for a step function load is called `isostatic response' and results in the `final state of isostatic equilibrium' which is hydrostatic unless an elastic lithosphere provides static support \citep{peltier1974,peltier1976,wu1982,han1995,cambiotti2013}.
This terminology, however, excludes from isostasy the intermediate stage of Airy isostatic balance illustrated in Figure~\ref{FigShapeGravTime}.
By contrast, the present work addresses long-term processes that have reached this particular stage where the shape ratio has become stationary.
As shown in Figure~\ref{FigShapeGravTime}, Airy isostatic balance, once achieved, is not very sensitive to the viscosity ratio between the sublayer and the shell as long as the shell has a higher viscosity.
The system can thus be modelled as a viscoelastic shell above an inviscid sublayer, as done by \citet{ermakov2017}.
In that case, the average viscosity of the shell does not affect Airy isostasy, but the variation of rheology with depth does (lateral variations are ignored for simplicity).
This assumption allows me to keep the model as close as possible to the traditional concept of isostasy, as well as to the elastic equilibrium approach.
This complements the work of \citet{zhong2000}, who studied the appearance of isostatic equilibrium in configurations where the layers are either viscous or elastic (see also \citet{zhong1997,zhong2002}), and the one of \citet{kamata2012}, who computed the long-term relaxation of stratified bodies with time-dependent viscosity.

Although viscous relaxation is an attractive mechanism to produce Airy isostasy, the lower viscosity at the bottom of the shell can lead to strong lateral flow which destroys the isostatic topography \citep{gratton1989,bird1991,mckenzie2000}.
The problem is particularly acute for icy satellites with subsurface oceans, because of the short relaxation time scale at the shell-ocean boundary which results from the low viscosity of ice at melting point and the free-slip condition at the interface.
Topography can instead be maintained in a dynamical equilibrium if nonuniform internal heating causes uneven melting/freezing at the bottom of the shell \citep{kamata2017,cadek2017,kvorka2018,cadek2019,cadek2019b}.
At long wavelengths, nonuniform heating is produced by tidal dissipation within the shell or the core.
If the satellite is at equilibrium, tidal dissipation is constant so that the melting/freezing rate is constant too.
In the Love number approach, this corresponds to a load increasing linearly with time, whose lateral flow results in a shell of constant shape.
Global equilibrium between tidally dissipated heat and heat loss through conduction or convection is a plausible outcome for a satellite in orbital resonance \citep{segatz1988,fuller2016,nimmo2018}.
However, the thermal-orbital evolution can also display oscillations around equilibrium \citep{ojakangas1986,fischer1990,hussmann2004,meyer2008}, in which case the load and the shape both depend on time.
For these reasons, I will examine Airy isostasy with variable loads, focussing on the case of constant shape, but considering also more general time-dependent loads.

In this paper, I extend the `Isostasy with Love' framework of Paper~I in order to describe Airy isostasy as resulting from viscoelastic relaxation.
The viscoelastic and viscous approaches are in principle equivalent in the long time limit, as there is no memory left of the initial elastic response.
While this equivalence seems natural in the time-dependent approach, in which the load is not modified after emplacement, it is less obvious in the stationary approach, in which the load evolves with time in order to maintain the boundary condition.
We thus have (too) many isostatic approaches based on sound physical principles: minimum elastic stress, viscoelastic evolution, and viscous evolution, with the two last ones in time-dependent and stationary versions.

\begin{figure}
\centering
    \includegraphics[width=6cm]{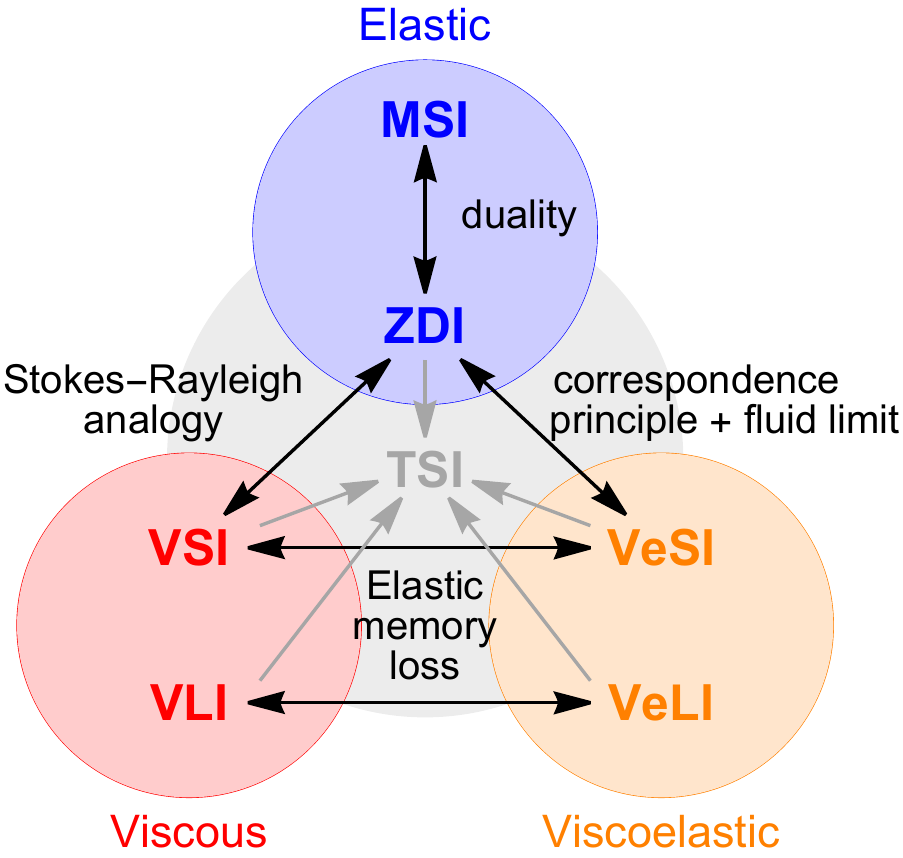}
   \caption[Relations between elastic, viscoelastic, and viscous isostatic models]{Relations between elastic, viscoelastic, and viscous isostatic models.
   The acronyms stand for minimum stress isostasy (MSI), zero deflection isostasy (ZDI), viscoelastic `constant shape' isostasy (VeSI), viscoelastic `constant load' isostasy (VeLI),
   viscous `constant shape' isostasy (VSI), viscous `constant load' isostasy (VLI).
    In the thin shell limit, all models tend to thin shell isostasy (TSI).
 }
   \label{FigIsoRelations}
\end{figure}

In Paper~I, I showed that the invariance of elastic isostasy under a global rescaling of the shear modulus means that it can be computed in the fluid limit.
This property also suggests a connection with the asymptotic state of a viscoelastic/viscous shell.
Similarly, viscoelastic isostasy is invariant under a global rescaling of viscosity, the counterpart of the fluid limit being the long-time limit.
If the shell is homogeneous, elastic and viscoelastic isostatic models actually depend on the same set of interior parameters.
Using the correspondence principle and the fluid limit, I will establish an exact relation between stationary viscoelastic isostasy and elastic isostasy; more general time-dependent loads will be shown to result in a similar Airy isostatic balance.
Besides, I will use the Stokes-Rayleigh analogy to relate stationary viscous isostasy to elastic isostasy.
By contrast, the time-dependent versions of viscoelastic and viscous isostasy are equivalent between them, but differ from the stationary version.
Fig.~\ref{FigIsoRelations} illustrates the relations between elastic, viscoelastic, and viscous approaches.
Existing isostatic models thus belong to two independent groups: the elastic/stationary approaches, and the time-dependent approaches.
In the thin shell limit, all models yield the same predictions which coincide with the thin shell model of \citet{dahlen1982}.
At high harmonic degree, shell thickness matters and the various models give slightly different predictions of geoid anomalies.

Viscoelastic Love numbers can be computed with standard methods \citep{sabadini2016} and various software packages \citep{spada2011}, although the latter should be modified to include forcing by internal loads.
Airy isostasy stricto sensu corresponds to the case of a shell of uniform rheology, but I will also discuss shells with depth-dependent rheology which are particularly relevant for icy moons with subsurface oceans \citep{cadek2019}.
Fully analytical solutions are feasible for simple configurations, and are given in the complementary software for a 3-layer incompressible body with homogeneous layers \citep{beuthe2020z}.

%%%%%%%%%%%%%%%%%%%%%%%%%%%%%%%%%%%%%%%%%%%%%%%%%%%%%%%%%%%%%%%%%%%%%%%%%%%%%%%%%%%%%%%%%%%%%%%%%%%%%%%
%%%%%%%%%%%%%%%%%%%%%%%%%%%%%%%%%%%%%%%%%%%%%%%%%%%%%%%%%%%%%%%%%%%%%%%%%%%%%%%%%%%%%%%%%%%%%%%%%%%%%%%
%%%%%%%%%%%%%%%%%%%%%%%%%%%%%%%%%%%%%%%%%%%%%%%%%%%%%%%%%%%%%%%%%%%%%%%%%%%%%%%%%%%%%%%%%%%%%%%%%%%%%%%

%%%%%%%%%%%%%%%%%%%%%%%%%%%%%%%%%%%%%%%%%%%%%%%%%%%%%%%%%%%%%%%%%%%%%%%%%%%%%%%%%%%%%%%%%%%%%%%%%%%%%%%
%%%%%%%%%%%%%%%%%%%%%%%%%%%%%%%%%%%%%%%%%%%%%%%%%%%%%%%%%%%%%%%%%%%%%%%%%%%%%%%%%%%%%%%%%%%%%%%%%%%%%%%
%%%%%%%%%%%%%%%%%%%%%%%%%%%%%%%%%%%%%%%%%%%%%%%%%%%%%%%%%%%%%%%%%%%%%%%%%%%%%%%%%%%%%%%%%%%%%%%%%%%%%%%
\section{Isostatic ratios in terms of deviatoric Love numbers}
\label{SectionLoading}

%%%%%%%%%%%%%%%%%%%%%%%%%%%%%%%%%%%%%%%%%%%%%%%%%%%%%%%%%%%%%%%%%%%%%%%%%%%%%%%%%%%%%%%%%%%%%%%%%%%%%%%
%%%%%%%%%%%%%%%%%%%%%%%%%%%%%%%%%%%%%%%%%%%%%%%%%%%%%%%%%%%%%%%%%%%%%%%%%%%%%%%%%%%%%%%%%%%%%%%%%%%%%%%
\subsection{Loading the shell}

\begin{figure}
\centering
   \includegraphics[width=\textwidth]{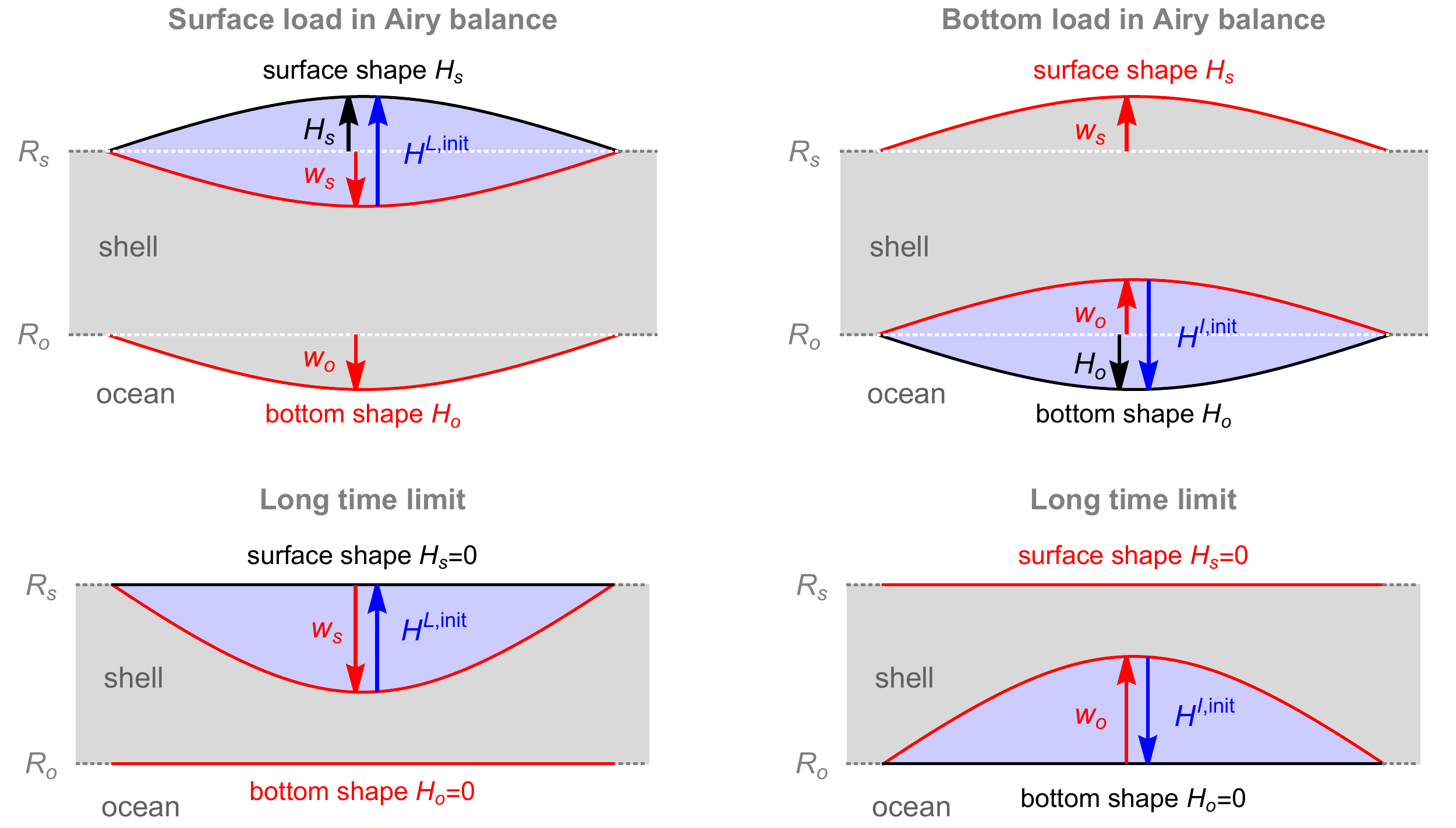}
   \caption[Loading of a viscoelastic shell]{
   Loading of a viscoelastic shell by a surface load (left panels) or a bottom load (right panels).
   The top row shows the stage of Airy isostatic balance while the bottom row shows the long time limit.
   In these examples, the shell is thickened by the load.
   The ocean density is twice the shell density.
}
   \label{FigDeformation2}
\end{figure}

In Paper~I, Airy isostasy was formulated as an elastic loading problem, more precisely as the mechanical equilibrium of loads applied at the surface and at the bottom of the shell.
The theoretical framework is the same in viscoelastic isostasy, except that the shell is now viscoelastic.
Besides, surface and internal loads do not need to be both present since mechanical balance results from viscoelastic relaxation.
This section summarizes formulas that are readily applicable to viscoelastic isostasy, whereas Appendix~\ref{AppendixElasticIsostasy} includes those that are specific to elastic isostasy.

The initial state, or unperturbed model, is spherically symmetric with a viscoelastic shell (or crust) floating on a fluid layer (ocean or asthenosphere) surrounding an elastic core (see Table~\ref{TableParamV}).
Surface and internal loads, respectively associated with the initial elevations of the surface ($H_n^{L,init}$) and shell-ocean boundary ($H_n^{I,init}$), are applied to the shell which responds by deforming and perturbing the initial gravity field (Fig.~\ref{FigDeformation2}).
The equations of equilibrium and Poisson's equation are linearized around the unperturbed state under the assumption that loads and deformations have a small amplitude.
Shape and gravitational perturbations are expanded in spherical harmonics with degree $n\geq2$.
At degree~1, the shape perturbation is fully compensated in the centre-of-mass frame (Section~3.3 of Paper~I).
By analogy with tidal forcing, the topographic loading is represented by the gravitational potentials produced by the initial elevations at the radii where they are located:
\begin{eqnarray}
H_n^{L,init} &=& \frac{1}{\xi_{sn}} \, \frac{U_n^L}{g_s} \, ,
\label{LoadPotential1} \\
H_n^{I,init} &=& \frac{1}{ x \Delta\xi_n } \, \frac{U_n^I}{g_s} \, ,
\label{LoadPotential2}
\end{eqnarray}
with parameters defined in Table~\ref{TableParamV}.
These formulas depend on the Airy assumption that the surface and internal loads are defined by undulations of the shell boundaries with respect to their unperturbed spherical state, so that the associated surface mass densities are $\rho_sH_n^{L,init}$ and $(\rho_o-\rho_s)H_n^{I,init}$, respectively.
According to potential theory (see Eqs.~(7)-(8) of Paper~I), the gravitational perturbations due to the combined loads at the shell boundaries are given by
\begin{eqnarray}
\Gamma_{sn}^{\rm direct} &=& U_n^L + x^{n+1} \, U_n^I \, ,
\\
\Gamma_{on}^{\rm direct} &=& x^n \, U_n^L + U_n^I \, .
\end{eqnarray}
The superscript `direct' denotes that the contribution due to the deformation of the body is not included.

%TABLE 1
\begin{table}[h]\centering
\ra{1.3}
\small
\caption[Internal structure parameters for a 3-layer body with homogeneous layers]{Internal structure parameters for a 3-layer body with homogeneous layers.}
\vspace{1.5mm}
\begin{tabular}{@{}lll@{}}
\toprule
Parameter &  Symbol & Nondimensional version \\
%\hline
\midrule
Surface (or shell) radius & $R_s$ & 1 \\
Ocean (or asthenosphere) radius & $R_o$ & $x=R_o/R_s$ \\
Shell thickness & $d_s$ & $\varepsilon=d_s/R_s=1-x$ \\
Bulk density & $\rho_b$ & 1 \\
Shell density & $\rho_s$ & $\xi_{sn}=\frac{3}{2n+1} \,  (\rho_s/\rho_b)$ \\
Ocean density & $\rho_o$ & $\xi_{on}=\frac{3}{2n+1} \,  (\rho_o/\rho_b)$ \\
Density contrast (ocean-shell) & $\rho_o-\rho_s$ & $\Delta\xi_n=\xi_{on}-\xi_{sn}$ \\
Surface gravity & $g_s$ & $\gamma_s=1$ \\
Gravity at bottom of shell & $g_o$ & $\gamma_o = g_o/g_s = (1+ (x^3-1) \, \xi_{s1} )/x^2$ \\
Elastic shear modulus (shell) & $\mu_{\rm s}$ & $\bar\mu_{\rm s}=\mu_{\rm s}/(\rho_b g_s R_s)$ \\
%\hline
\addlinespace[2pt]
\bottomrule
\end{tabular}
\label{TableParamV}
\end{table}%

Radial Love numbers ($h_j^L,h_j^I$) relate the radial deformation of the shell boundaries (excluding the load, as in Fig.~\ref{FigDeformation2}) to the load potentials:
\begin{equation}
w_{jn} = \frac{1}{g_s} \left( h_j^L \, U_{n}^L+ h_j^I  \, U_{n}^I \right) .
\label{wn}
\end{equation}
The indices $j=(s,o)$ label the top and bottom of the shell ($s$ for surface and $o$ for ocean).
Gravitational Love numbers ($k_j^L,k_j^I$) do the same job for the gravitational perturbation induced by the deformation of the body:
\begin{equation}
\Gamma_{jn}^{\rm induced} = k_j^L \, U_{n}^L + k_j^I \, U_{n}^I \, .
\end{equation}
However, we are neither interested by the deformation nor by the induced gravity, which are not observable, but instead by the shape of the shell boundaries $(H_{sn},H_{on})$ and the total gravitational perturbation at the same interfaces $(\Gamma_{sn},\Gamma_{on})$.
The former are the sum of the initial elevation (`direct effect') and the radial deformation, while the latter are the sum of direct and induced gravitational contributions.
The direct effects can be written in terms of the negative of the \textit{fluid Love numbers} ($h_j^{J \circ},k_j^{J\circ}$) given in Table~\ref{TableLoveV}, so that the shape and total gravitational perturbation read
\begin{eqnarray}
H_{jn} &=& \left( - h_j^{L \circ} + h_j^L  \right) \frac{U_{n}^L}{g_s} + \left( -h_j^{I \circ} + h_j^I  \right) \frac{U_n^I}{g_s} \, ,
\label{ShapeDev0} \\
\Gamma_{jn} &=& \left( - k_j^{L \circ}  + k_j^L \right) U_{n}^L + \left( - k_j^{I \circ}  + k_j^I \right) U_{n}^I \, .
\label{GravDev0}
\end{eqnarray}
Fluid Love numbers were computed in Paper~I from the elastic-gravitational equations; they are not sensitive to the deep interior (below the shell).
If there is no elastic lithosphere, the shell tends in the long time limit to a fluid state and Love numbers tend to their fluid values, so that the shape of the shell boundaries and the total gravitational perturbation both vanish (Fig.~\ref{FigDeformation2}).
The vanishing of the shape depends on the Airy assumption that loads are defined by undulations of the shell boundaries (Eqs.~(\ref{LoadPotential1})-(\ref{LoadPotential2})).

It is convenient to define \textit{deviatoric Love numbers} which vanish in the long time (or fluid) limit:
\begin{eqnarray}
\hat h_j^J =  h_j^J - h_j^{J \circ}  \, ,
 \\
\hat k_j^J =  k_j^J - k_j^{J \circ} \, ,
\label{LoveDecomp}
\end{eqnarray}
where $j=(s,o)$ and $J=(L,I)$.
The shape and total gravitational perturbation can then be written in terms of load potentials and deviatoric Love numbers:
\begin{eqnarray}
H_{jn} &=& \frac{1}{g_s} \left( \hat h_j^L \, U_{n}^L + \hat h_j^I \, U_n^I \right) ,
\label{ShapeDev} \\
\Gamma_{jn} &=& \hat k_j^L \, U_{n}^L + \hat k_j^I \, U_{n}^I \, .
\label{GravDev}
\end{eqnarray}
These formulas make it explicit that the shape and the total gravitational perturbation vanish in the long time (or fluid) limit, together with the deviatoric Love numbers.

%TABLE 2
\begin{table}[h]\centering
\ra{1.3}
\small
\caption[Love numbers as a sum of fluid and deviatoric components]{Radial Love numbers $h_j^J$ and gravitational Love numbers $k_j^J$ (implicitly at harmonic degree $n$) as a sum of fluid and deviatoric (`Dev') components.
The superscript denotes the forcing ($L$ for surface load; $I$ for internal load); the subscript denotes the interface where the Love number is evaluated ($s$ for top of shell; $o$ for bottom of shell).
}
\vspace{1.5mm}
\begin{tabular}{@{}llllll@{}}
\toprule
\multicolumn{3}{c}{Radial} & \multicolumn{3}{c}{Gravitational} \\
\cmidrule(lr){1-3} \cmidrule(lr){4-6}
Full & \hspace{1.5mm} Fluid & Dev. & Full  & \hspace{1.5mm} Fluid & Dev. \\
$h_j^J$ & \hspace{1.5mm}  $h_j^{J\circ}$ & $\hat h_j^J$ & $k_j^J$ & \hspace{1.5mm} $k_j^{J\circ}$ & $\hat k_j^J$ \\
\midrule
$h_s^L$ & $-1/\xi_{sn}$ & $\hat h_s^L$ & $k_s^L$ & $-1$ & $\hat k_s^L$ \\ 
$h_o^L$ & \hspace{1.5mm} $0$ & $\hat h_o^L$ & $k_o^L$ & $-x^n$ & $\hat k_o^L$ \\ 
$h_s^I$ & \hspace{1.5mm} $0$ & $\hat h_s^I$ & $k_s^I$ &  $-x^{n+1}$ & $\hat k_s^I$ \\ 
$h_o^I$ &  $-1/(x \Delta\xi_n)$ & $\hat h_o^I$ & $k_o^I$ &  $-1$ & $\hat k_o^I$ \\ 
\addlinespace[2pt]
\bottomrule
\end{tabular}
\label{TableLoveV}
\end{table}%

%%%%%%%%%%%%%%%%%%%%%%%%%%%%%%%%%%%%%%%%%%%%%%%%%%%%%%%%%%%%%%%%%%%%%%%%%%%%%%%%%%%%%%%%%%%%%%%%%%%%%%%
%%%%%%%%%%%%%%%%%%%%%%%%%%%%%%%%%%%%%%%%%%%%%%%%%%%%%%%%%%%%%%%%%%%%%%%%%%%%%%%%%%%%%%%%%%%%%%%%%%%%%%%
\subsection{Isostatic ratios}

\textit{Isostatic ratios} are nondimensional quantities characterizing the output of an isostatic model.
They depend on deviatoric Love numbers and on the \textit{loading ratio} which is the ratio of the internal load to the surface load:
\begin{equation}
\zeta_n = U_n^I/U_n^L \, .
\label{LoadingRatio}
\end{equation}
In elastic isostasy, the loading ratio is determined by the choice of the isostatic prescription (e.g.\ minimum stress), but it is a free parameter in viscoelastic isostasy, for which arbitrary loads can be applied at the surface and at the bottom of the shell.
It is not a priori clear whether viscoelastic isostasy depends or not on the loading ratio.

Shape and gravitational perturbations (Eqs.~(\ref{ShapeDev})-(\ref{GravDev})) can be combined into various nondimensional isostatic ratios, the most important of which is the \textit{compensation factor} $F_n$, defined as the ratio of the geoid perturbation $\Gamma_{sn}/g_s$ to the surface shape $H_{sn}$, normalized so that $F_n$ varies between 0 (full compensation) and 1 (no compensation):
\begin{equation}
F_n \,=\, \frac{1}{\xi_{sn}} \, \frac{\Gamma_{sn}}{g_sH_{sn}}
\,=\, \frac{1}{\xi_{sn}} \, \frac{ \hat k_s^L + \zeta_n \, \hat k_s^I }{ \hat h_s^L + \zeta_n \, \hat h_s^I } \, .
\label{Fn}
\end{equation}
Its importance comes from the fact that it can be computed from quantities observable at the surface. 
By contrast, the \textit{shape ratio} $S_n$ and the \textit{topographic ratio} $T_n$ are not easily observable, but are useful proxies when constructing new isostatic models:
\begin{eqnarray}
S_n &=& \frac{H_{on}}{H_{sn}}
\,\, = \,\, \frac{ \hat h_o^L + \zeta_n \, \hat h_o^I}{\hat h_s^L + \zeta_n \, \hat h_s^I} \, ,
\label{Sn} \\
T_n &=& \frac{H_{on}-\Gamma_{on}/g_o}{H_{sn}-\Gamma_{sn}/g_s}
\,\, = \,\, \frac{1}{\gamma_o} \, \frac{ \hat t^{\, L}_o + \zeta_n \, \hat t^{\, I}_o}{ \hat t^{\, L}_s + \zeta_n \, \hat t^{\,I} _s} \, ,
\label{Tn}
\end{eqnarray}
where $\hat  t^{\, J}_j = \gamma_j \, \hat  h^{J}_j - \hat  k^{J}_j$.
If the body is incompressible and the density of the shell is uniform, gravitational  Love numbers can be computed in terms of radial Love numbers at the shell boundaries.
In that case, isostatic ratios are related by
\begin{eqnarray}
S_n &=& \frac{ (1-a) \left(\gamma_o T_n \right) +  c }{ b \left(\gamma_o T_n \right) + d }  \, ,
\label{TnSnRelation} \\
\xi_{sn} F_n &=& b \, S_n + a \, ,
\label{FnSnRelation}
\end{eqnarray}
with $(a,b,c,d)$ given in Table~3 of Paper~I.

%%%%%%%%%%%%%%%%%%%%%%%%%%%%%%%%%%%%%%%%%%%%%%%%%%%%%%%%%%%%%%%%%%%%%%%%%%%%%%%%%%%%%%%%%%%%%%%%%%%%%%%
%%%%%%%%%%%%%%%%%%%%%%%%%%%%%%%%%%%%%%%%%%%%%%%%%%%%%%%%%%%%%%%%%%%%%%%%%%%%%%%%%%%%%%%%%%%%%%%%%%%%%%%
%%%%%%%%%%%%%%%%%%%%%%%%%%%%%%%%%%%%%%%%%%%%%%%%%%%%%%%%%%%%%%%%%%%%%%%%%%%%%%%%%%%%%%%%%%%%%%%%%%%%%%%
\section{Viscoelastic isostasy}
\label{SectionViscoelastic}

%%%%%%%%%%%%%%%%%%%%%%%%%%%%%%%%%%%%%%%%%%%%%%%%%%%%%%%%%%%%%%%%%%%%%%%%%%%%%%%%%%%%%%%%%%%%%%%%%%%%%%%
%%%%%%%%%%%%%%%%%%%%%%%%%%%%%%%%%%%%%%%%%%%%%%%%%%%%%%%%%%%%%%%%%%%%%%%%%%%%%%%%%%%%%%%%%%%%%%%%%%%%%%%
\subsection{Time-dependent Love numbers}

\begin{figure}
   \centering
    \includegraphics[width=\textwidth]{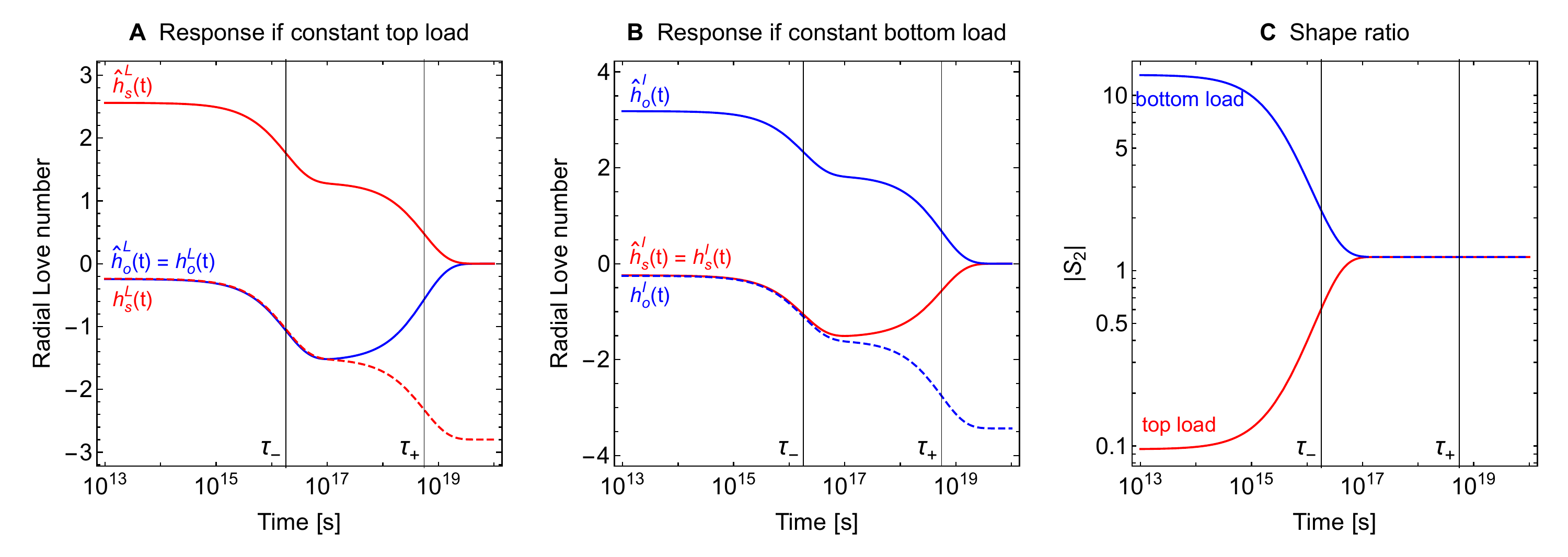}
   \caption[Radial displacement of shell boundaries under a constant degree-two load]{Radial displacement of shell boundaries under a constant degree-two load, as a function of time: (A) top load; (B) bottom load; (C) absolute value of shape ratio.
    This example is based on the two-layer Ceres model of \citet{ermakov2017} (see Fig.~\ref{FigShapeGravTime}).
    The deformation of the top and bottom shapes (including the load) is quantified by the deviatoric Love number (solid curves), which tends to zero in the long time limit.
    The deformation of the interface between shell and load is quantified by the full Love number (dashed curves).
    Red and blue curves correspond to the top and bottom of the shell, respectively.
    Vertical lines indicate the two viscoelastic decay times ($\tau_-=1.80\times10^{16}\rm\,s$ and $\tau_+=5.52\times10^{18}\rm\,s$).
   }
   \label{FigLoveTime}
\end{figure}

In the viscoelastic approach, the shell initially at hydrostatic equilibrium is loaded at time $t=0$ by surface and bottom loads with some constant loading ratio $\zeta^V_n$.
If the viscosity of the shell is not infinite, the system evolves with time towards hydrostatic equilibrium.
As the shell flows away under the loads, the internal stresses decrease and the shell finds itself in a state of approximate equilibrium, or isostatic balance: the shapes of the top and bottom shell boundaries tend to zero, but their ratio is constant (Fig.~\ref{FigLoveTime}).
The shape ratio and compensation factors are thus defined for viscoelastic isostasy by the long-time limit of Eqs.~(\ref{Fn})-(\ref{Sn}) in which Love numbers depend on time:
\begin{eqnarray}
F_n &=& \lim_{t\rightarrow\infty} \frac{1}{\xi_{sn}} \, \frac{ \hat k_s^L(t) + \zeta^V_n \, \hat k_s^I(t) }{ \hat h_s^L(t) + \zeta^V_n \, \hat h_s^I(t) } \, ,
\label{CompensRatioV} \\
S_n &=& \lim_{t\rightarrow\infty} \frac{ \hat h_o^L(t) + \zeta^V_n \, \hat h_o^I(t)}{\hat h_s^L(t) + \zeta^V_n \, \hat h_s^I(t)} \, ,
\label{ShapeRatioV}
\end{eqnarray}
where $\zeta^V_n$ denotes the loading ratio in viscoelastic isostasy.
If the load evolves in time, its time-dependence is included in the time-dependent Love numbers (see Eq.~(\ref{finalvalue}) and Section~\ref{ViscoelasticConstantShape}).
Recall that the topographic ratio and compensation factor can be computed from the shape ratio if the body is incompressible and the shell of uniform density (Eqs.~(\ref{TnSnRelation})-(\ref{FnSnRelation})).

%%%%%%%%%%%%%%%%%%%%%%%%%%%%%%%%%%%%%%%%%%%%%%%%%%%%%%%%%%%%%%%%%%%%%%%%%%%%%%%%%%%%%%%%%%%%%%%%%%%%%%%
%%%%%%%%%%%%%%%%%%%%%%%%%%%%%%%%%%%%%%%%%%%%%%%%%%%%%%%%%%%%%%%%%%%%%%%%%%%%%%%%%%%%%%%%%%%%%%%%%%%%%%%
\subsection{Correspondence principle}

The principle of correspondence \citep{peltier1974} gives a practical procedure to compute time-dependent Love numbers of a viscoelastic body:
\begin{enumerate}
\item compute the solution of the elastic problem;
\item replace the elastic shear modulus $\mu_{\rm e}$ by the $s$-dependent function $\tilde\mu(s)$ characterizing the rheology, $s$ being the Laplace domain variable;
\item multiply the result by the Laplace transform $\tilde T(s)$ of the time history of the load;
\item take the inverse Laplace transform of this transformed solution (see Table~\ref{TableLaplace}).
\end{enumerate}
For Maxwell rheology, the shear modulus in the Laplace domain reads
\begin{equation}
\tilde\mu(s) = \frac{\mu_{\rm e} \, s}{s+\mu_{\rm e}/\eta} \, ,
\label{maxwell}
\end{equation}
where $\mu_{\rm e}$ is the elastic shear modulus and $\eta$ is the viscosity (see \citet{peltier1982} for other rheological models).
In the large-$s$ limit, the material is purely elastic;
in the small-$s$ limit, it looks like a fluid with an asymptotic behaviour governed by the viscosity:
\begin{eqnarray}
\lim_{s\rightarrow\infty} \tilde \mu(s) &=& \mu_{\rm e} \, ,
\\
\lim_{s\rightarrow0} \, \tilde \mu(s) &=& 0 \, ,
\\
\partial_s \tilde \mu(s)  \Big|_{s=0} &=& \eta \, .
\label{muderivative}
\end{eqnarray}

I will now explain how viscoelastic Love numbers are related to their elastic, fluid, and deviatoric components.
Equations are given below for radial Love numbers, but similar equations hold for gravitational Love numbers. 
In the Laplace domain (in which functions are denoted with a tilde), Love numbers can be written as a sum over normal modes of viscous gravitational relaxation \citep{peltier1976,wu1982,sabadini2016}.
\begin{equation}
\tilde h_j^J(s) = h_{j\rm e}^J + \sum_p \frac{r^J_{jp}}{s - s_p} \, ,
\label{LoveLaplace}
\end{equation}
where $h_{j\rm e}^J$ is the elastic Love number (it does not depend on $s$) and $r^J_{jp}$ are the residues associated with the poles $s=s_p$ which are on the negative real  axis ($s_p<0$).
Since the poles are the eigenvalues of the homogeneous viscoelastic-gravitational problem, they do not depend on the type of loading and are also poles for the gravitational Love numbers.
The simplest interior model appropriate for isostasy (incompressible body with homogeneous shell, inviscid ocean, and elastic core) has two modes because there is one viscoelastic layer (the shell) which has two boundaries with a density contrast (see Section 1.8 of \citet{sabadini2016}).

In the large-$s$ limit, Laplace-domain Love numbers tend to their elastic values:
\begin{equation}
\lim_{s\rightarrow\infty} \tilde h_j^J(s) = h_{j\rm e}^J \, .
\end{equation}
In the small-$s$ limit, they tend to their fluid values (Table~\ref{TableLoveV}):
\begin{eqnarray}
\lim_{s\rightarrow0} \tilde h_j^J(s)
&=& h_{j\rm e}^J - \sum_p \frac{r^J_{jp}}{s_p}
\nonumber \\
&=& h_j^{J\circ} \, .
\end{eqnarray}
The remainder is the deviatoric component which vanishes in the small-$s$ limit:
\begin{eqnarray}
\hat{\tilde h}_j^J(s)
&=&  \tilde h_j^J(s) - h_j^{J\circ}
\nonumber \\
&=& \sum_p \frac{r^J_{jp}}{s_p} \, \frac{s}{s - s_p} \, .
\label{hdevLaplace}
\end{eqnarray}
In analogy with Eq.~(\ref{LovePartialE}), the partial derivative of the Laplace-domain Love number with respect to $s$ and evaluated at $s=0$ is denoted
\begin{eqnarray}
\dot {\tilde h}_j^J
&=& \partial_s \tilde h_j^J(s) \Big|_{s=0}
\nonumber \\
&=&  - \sum_p \frac{r^J_{jp}}{(s_p)^2} \, .
\label{LovePartialV}
\end{eqnarray}

In order to go over to the time domain, I must specify the time history of the load.
Given the Laplace transform $\tilde T(s)$ of the loading time history, the final value theorem \citep{peltier1974,dyke2014} relates the long-time limit of the time-dependent Love number to the small-$s$ limit of its Laplace transform:
\begin{equation}
\lim_{t\rightarrow\infty} \hat h_j^J(t) = \lim_{s\rightarrow0} s \left( \tilde T(s) \, \hat{\tilde h}_j^J(s) \right) .
\label{finalvalue}
\end{equation}

%TABLE 1
\begin{table}[h]\centering
\ra{1.3}
\small
\caption[Laplace transforms]{Laplace transforms used in this paper.
The first two columns give the time dependence of the load and its Laplace transform; the last two columns give the inverse Laplace transform of the product of the load with the function $f(s)=s/(s-a)$ ($a$ is real and negative) and its asymptotic behaviour at large times.
$u(t)$ denotes the unit step function; $q$ is a positive integer;
$\Gamma(q)$ and $\Gamma(q,a)$ are the gamma function and the incomplete gamma function, respectively.}
\vspace{1.5mm}
\begin{tabular}{@{}llll@{}}
\toprule
%$\tilde T(s)$ &  $T(t)$ & ${\cal L}^{-1}(\tilde T(s) \, \frac{s}{s-a} )$ \\
$T(t)$ & $\tilde T(s)$ &   ${\cal L}^{-1}(\tilde T(s) \, f(s) )$ & $t\rightarrow\infty$ \\
\midrule
%$1$ & $1/s$ &  $\exp(at)$ & $\exp(at)$ \\
%$u(t)$ & $1/s$ &  $e^{at} \, u(t)$ & $e^{at}$ \\
$u(t)$ & $1/s$ &  $\exp(at) \, u(t)$ & $\exp(at)$ \\
%$u(t-t_1)$ & $e^{-t_1 s}/s$ &  $e^{a(t-t_1)} \, u(t-t_1)$ & $e^{a(t-t_1)}$ \\
$u(t-t_1)$ & $\exp(-t_1 s)/s$ &  $\exp(a(t-t_1)) \, u(t-t_1)$ & $\exp(a(t-t_1))$ \\
$t \, u(t)$ & $1/s^2$ & $a^{-1} (\exp(at)-1) u(t)$ & $-1/a$ \\
$t^q \, u(t)$ & $\Gamma(q+1)/s^{q+1}$ &  $q \, a^{-q} \exp(at) (\Gamma(q)-\Gamma(q,a t)) \, u(t)$ & $(-q/a) \, t^{q-1}$ \\
$\cos(\omega t) u(t)$ & $s/(s^2+\omega^2)$ & $\frac{a^2\exp(at)+\omega^2\cos(\omega t)+a\omega\sin(\omega t)}{a^2+\omega^2} \, u(t)$ & $\frac{\omega^2\cos(\omega t)+a\omega\sin(\omega t)}{a^2+\omega^2}$ \\
%\hline
\addlinespace[2pt]
\bottomrule
\end{tabular}
\label{TableLaplace}
\end{table}%

%%%%%%%%%%%%%%%%%%%%%%%%%%%%%%%%%%%%%%%%%%%%%%%%%%%%%%%%%%%%%%%%%%%%%%%%%%%%%%%%%%%%%%%%%%%%%%%%%%%%%%%
%%%%%%%%%%%%%%%%%%%%%%%%%%%%%%%%%%%%%%%%%%%%%%%%%%%%%%%%%%%%%%%%%%%%%%%%%%%%%%%%%%%%%%%%%%%%%%%%%%%%%%%
\subsection{Constant load}
\label{ViscoelasticConstantLoad}

Constant loading means that the load applied at $t=0$ remains constant beyond some threshold $t_0>0$.
This model is relevant to the relaxation of topography that is not modified after its emplacement, for example small-scale features such as craters and dormant volcanoes, or large-scale loads such as the fossil figure due to faster rotation in the past.
Since any continuous loading history can be approximated by a series of discrete steps \citep{peltier1976},
it is usual to start with the simpler case of a surface (or internal) load brought from infinity at time $t=0$ and remaining constant for $t>0$.
The time history of this load is represented by the Heaviside (or unit step) function $u(t)$, the Laplace transform of which is $\tilde T(s)=1/s$ (see Table~\ref{TableLaplace}).
The final value theorem implies that the long time limit of the deviatoric Love number is equal to the small-$s$ limit of the Laplace-domain deviatoric Love number, which vanishes:
\begin{equation}
 \lim_{t\rightarrow\infty} \hat h_j^{J}(t) \Big|_{\rm VeLI} = \hat{\tilde h}_j^J(0) \, = \, 0 \, ,
\end{equation}
where `VeLI' stands for \textbf{V}isco-\textbf{e}lastic (constant) \textbf{L}oad \textbf{I}sostasy.
This equation is not sufficient to compute the shape ratio and compensation factor which have the indeterminate form $0/0$.
One needs to know precisely how the deviatoric Love number decreases as $s$ tends to zero.
For this purpose, I compute the inverse Laplace transform of the product of Eq.~(\ref{hdevLaplace}) and $\tilde T(s)=1/s$ (see Table~\ref{TableLaplace}): 
\begin{equation}
\hat h_j^{J}(t) \Big|_{\rm VeLI} = \sum_p \frac{r^J_{jp}}{s_p} \,  \exp(s_p \, t) \, u(t) \, .
\label{hdevH}
\end{equation}
In order to facilitate comparisons with \citet{peltier1976}, I also give the time evolution of the full Love number (for $t>0$):
\begin{equation}
h_j^{J}(t) \Big|_{\rm VeLI} = h_{j\rm e}^J + \sum_p \frac{r^J_{jp}}{s_p} \, \Big( \exp(s_p \, t) -1 \Big) .
\end{equation}

In the long-time limit, the time evolution of the Love number is given by the exponential term with the longest time scale: $\tau_+=-1/s_+$, where $s_+$ denotes the pole closest to zero.
The shape ratio is thus equal to the ratio of the long time-scale residues:
\begin{equation}
S_n^{\rm VeLI} = \frac{ r^L_{o+} + \zeta^V_n \, r^I_{o+}}{r^L_{s+} + \zeta^V_n \, r^I_{s+}} \, .
\end{equation}
The VeLI shape ratio does not depend on the loading ratio because
\begin{equation}
S_n^{\rm VeLI} = \frac{ r^L_{o+}}{r^L_{s+}} = \frac{r^I_{o+}}{r^I_{s+}}\, .
\label{SnVeLI}
\end{equation}
This property is implied by the $\mu$-invariance of elastic isostasy (see Eq.~(\ref{IdentityConstantLoad})).

The compensation factor is computed in the same way.
The result can be written as
\begin{equation}
F_n^{\rm VeLI} = \frac{1}{\xi_{sn}} \, \frac{ r'^L_{s+} + \zeta^V_n \, r'^I_{s+}}{r^L_{s+} + \zeta^V_n \, r^I_{s+}} \, ,
\label{FnVeLI}
\end{equation}
where $r'^J_{j+}$ are the long time-scale residues for gravitational Love numbers.
It is simpler, however, to compute $F_n$ from $S_n$ with Eq.~(\ref{FnSnRelation}) if the body is incompressible and the shell is of uniform density.
As for the shape ratio, the VeLI compensation factor does not depend on the loading ratio.

More complicated loading histories (but constant for $t>t_0$) can be approximated by a series of unit steps starting at different times $0<t_1<t_2<...<t_0$.
A unit step load shifted at time $t=t_1$ has the effect of translating the result in the time domain (see Table~\ref{TableLaplace}), so that Eq.~(\ref{hdevH}) is multiplied by $\exp(-s_p t_1)$.
Such factors drop out of VeLI isostatic ratios, which are thus unaffected by the non-constant loading prior to the threshold $t=t_0$.
As expected, the isostatic balance keeps no memory of its early loading history.

Suppose now that the body is incompressible, the shell is homogenous, and the core is elastic (including the limiting vases of a fluid or an infinitely rigid core).
In that case, there are only two relaxation modes (see Appendix~\ref{AppendixA1}): Love numbers evolve in time as superpositions of a short-time decay mode, corresponding to the shell boundaries moving in the same direction, and a long-time decay mode, corresponding to shell boundaries moving in opposite directions, as shown by the solid curves in Fig.~\ref{FigLoveTime} (the same feature occurs in the viscous model, see \citet{hager1979}).
Assuming that the ocean and the core are homogeneous, I derive in  Appendix~\ref{AppendixPolesResidues} analytical formulas for the decay times and the shape ratio in terms of generic Love number coefficients.
Among several equivalent expressions (Eqs.~(\ref{SnVeLItop})-(\ref{SnVeLIAnalyticalBot})), the shape ratio is for example given by
\begin{equation}
S_n^{\rm VeLI} = \frac{F \mu_+ - \xi_{sn} \, B_s^L}{x\Delta\xi_n \, B_s^I } \, ,
\label{SnVeLIAnalytical}
\end{equation}
where the Love number coefficients $(B_s^L,B_s^I,F)$ and the pole $\mu_+$ are defined by Eqs.~(\ref{hgenV}) and (\ref{mupm}).
The analytical formulas are coded in the complementary software.

The VeLI isostatic ratios have the following general properties:
\begin{enumerate}
\item 
They are independent of the position of the load (at surface or at shell-ocean interface); more generally, they are independent of the loading ratio (Eq.~(\ref{SnVeLI})).
\item
They do not depend on the elastic properties of the shell (elastic memory loss).
They should thus be identical to the isostatic ratios of viscous isostasy with constant load (see Section~\ref{SectionViscousConstantLoad}).
\item 
They do not change if the viscosity of the shell is rescaled by a global factor ($\nu$-invariance).
They are thus independent of the viscosity of the shell if it is uniform (see Eqs.~(\ref{SnVeLItop})-(\ref{SnVeLIbot}))
\item 
They have the same thin shell limit as elastic isostasy (Eqs.~(\ref{ShapeRatioTSI}) and (\ref{SnVeLIthinshell1})-(\ref{SnVeLIthinshell2}))).
\item
At high harmonic degree, the shape ratio for a homogeneous shell diverges to $-\infty$ if $\psi = \gamma_o \, x\Delta\xi_1/\xi_{s1}\leq1$ (the most physically plausible case); it tends to zero if $\psi>1$ (Eqs.~(\ref{SnVeLIAsymptoticTop})-(\ref{SnVeLIAsymptoticBot})).
The compensation factor for a homogeneous shell does not diverge.
\end{enumerate}
The viscoelastic-viscous equivalence does not extend to decay times which are somewhat affected by the elastic shear modulus, but the discrepancy is small for the longest decay time (Eqs.~(\ref{DecayTimesViscoelastic})-(\ref{DecayTimesEquiv})).
In the thin shell limit, the longest relaxation time ($\tau_+=-1/s_+$) can be approximated from Eqs.~(\ref{muplusTS}) and (\ref{splusapprox}),
\begin{equation}
\tau_+ \cong \frac{\eta}{\rho_sg_sd} \, \frac{2\left(2n^2+2n-1\right)}{ n \left(n+1\right)} \, \frac{\xi_{o1}}{\Delta\xi_1} \, ,
\end{equation}
which agrees with the prediction of the viscous flow model in Cartesian geometry, taken in the thin shell limit: $\tau_+\cong(4\eta/\rho_sg_sd)(\rho_o/(\rho_o-\rho_s)$ (see Eq.~(A33) of \citet{solomon1982}).

%%%%%%%%%%%%%%%%%%%%%%%%%%%%%%%%%%%%%%%%%%%%%%%%%%%%%%%%%%%%%%%%%%%%%%%%%%%%%%%%%%%%%%%%%%%%%%%%%%%%%%%
%%%%%%%%%%%%%%%%%%%%%%%%%%%%%%%%%%%%%%%%%%%%%%%%%%%%%%%%%%%%%%%%%%%%%%%%%%%%%%%%%%%%%%%%%%%%%%%%%%%%%%%
\subsection{Constant shape}
\label{ViscoelasticConstantShape}

Consider now loading with constant shape, either at the surface or at the shell-ocean interface.
In the long-time limit, the final value theorem (Eq.~(\ref{finalvalue})) shows that the time-dependent deviatoric Love number tends to a non-zero constant if $\tilde T(s)=1/s^2$, which corresponds to a load increasing linearly with time (see Table~\ref{TableLaplace}).
For example, the melting (of the shell) or freezing (of the ocean) due to a constant heat flux from below can be interpreted as a bottom load increasing linearly with time.
For `constant shape' loading, the long-time limit of the deviatoric Love number is thus given by
\begin{equation}
\lim_{t\rightarrow\infty} \hat h_j^{J}(t) \Big|_{\rm VeSI}
=   - \sum_p \frac{r^J_{jp}}{(s_p)^2}
\,\, = \,\,  \dot {\tilde h}_j^J \, ,
\label{hdevC}
\end{equation}
where the last equality results from Eq.~(\ref{LovePartialV}) and `VeSI' stands for \textbf{V}isco-\textbf{e}lastic (constant) \textbf{S}hape \textbf{I}sostasy. 
The shape ratio is thus given by a formula very similar to the one obtained in the fluid limit of elastic isostasy (Eq.~(\ref{ShapeRatio0})), except that partial derivatives are evaluated in the Laplace domain:
\begin{eqnarray}
S_n^{\rm VeSI} &=& \frac{ \dot {\tilde h}_o^L + \zeta^V_n \, \dot {\tilde h}_o^I}{\dot {\tilde h}_s^L + \zeta^V_n \, \dot {\tilde h}_s^I} \, .
\label{ShapeRatioVC}
\end{eqnarray}
Suppose now that the shell has a Maxwell rheology (or another rheology which is fluid-like in the long term) and is stratified into $N$ layers with viscosities $(\eta_1,...,\eta_N)$.
Using the correspondence principle and the chain rule, I relate the partial derivative of the Laplace-domain Love number to the partial derivative of the Love number of an elastic body with a shell stratified into $N$ layers with shear moduli $(\mu_1,...,\mu_N)$:
\begin{eqnarray}
\dot {\tilde h}_j^J
&=& \sum_{i=1}^N \left. \left( \frac{\partial}{\partial \mu_i} \, h_{j\rm e}^J \right) \left( \frac{\partial}{\partial s} \,  \tilde \mu_i(s) \right) \right|_{s=0}
\nonumber \\
&=& \eta_0 \sum_{i=1}^N \left. \frac{\eta_i}{\eta_0} \, \frac{\partial}{\partial \mu_i} \, h_{j\rm e}^J \right|_{\mu_1=0 ... \mu_N=0} \, .
\label{htildedot}
\end{eqnarray}
The elastic shear moduli of the viscoelastic model play no role in this expression because Eq.~(\ref{muderivative}) depends only on the viscosity.
Eq.~(\ref{htildedot}) looks very much like the partial derivative of the elastic Love number in the fluid limit (Eq.~(\ref{LovePartialDiscrete})).
It can indeed be written as
\begin{equation}
\dot {\tilde h}_j^J = \eta_0 \, \dot h_j^J \, ,
\end{equation}
if the elastic shell in the RHS has a depth-dependent elastic shear modulus varying in the same way as the viscosity of the viscoelastic model in the LHS:
\begin{equation}
\frac{\eta_i}{\eta_0} = \frac{\mu_i}{\mu_0} \, .
\end{equation}
In that case, the formulas for the shape ratio and compensation factor have the same form as those of elastic isostasy in the fluid limit (Eq.~(\ref{ShapeRatio0})) with the substitution $\zeta_n^\circ\rightarrow\zeta_n^V$:
\begin{eqnarray}
S_n^{\rm VeSI} &=& \frac{ \dot h_o^L + \zeta^{\rm V}_n \, \dot h_o^I}{\dot h_s^L + \zeta^{\rm V}_n \, \dot h_s^I} \, .
\label{ShapeRatioVC0}
\end{eqnarray}
Thus, `constant shape' viscoelastic isostasy gives the same result as zero-deflection isostasy (ZDI) if $\zeta^{\rm V}_n$ coincides with the corresponding fluid loading ratio (Eq.~(\ref{zetaZDI0})):
\begin{equation}
\zeta^{\rm V}_n = - \frac{\Delta\xi_1}{\xi_{s1}} \, \frac{x}{\alpha} \, .
\label{EquivZetaAlpha}
\end{equation}
Therefore, `constant shape' viscoelastic isostasy with purely surface loading ($\zeta^{\rm V}_n=0$) is equivalent to elastic isostasy with zero deflection at the shell-ocean boundary ($\alpha=\infty$).
Conversely, `constant shape' viscoelastic isostasy with purely bottom loading  ($\zeta^{\rm V}_n=\pm\infty$) is equivalent to elastic isostasy with zero deflection at the surface ($\alpha=0$).
If the body is incompressible and has a homogeneous shell, the shape ratio can be expressed in terms of generic coefficients of Love numbers, as in Eq.~(\ref{ShapeRatioZDIGen}):
\begin{equation}
S_n^{\rm VeSI} = - \frac{\xi_{s1}}{x\Delta\xi_1} \, \frac{B_o^L - (\xi_{s1}/x\Delta\xi_1) \, \zeta^{\rm V}_n \, B_s^L}{B_o^I - (\xi_{s1}/x\Delta\xi_1) \, \zeta^{\rm V}_n \, B_s^I} \, .
\label{ShapeRatioVeSIGen}
\end{equation}
This formula is coded in the complementary software.

The VeSI isostatic ratios have the following properties:
\begin{enumerate}
\item 
They depend on the position of the load (at surface or at shell-ocean interface); more generally, they depend on the loading ratio $\zeta^{\rm V}_n$.
\item
They do not depend on the elastic properties of the shell (elastic memory loss).
They should thus be identical to the isostatic ratios of viscous isostasy with constant shape (see Section~\ref{SectionViscousConstantShape}).
\item 
They are $\nu$-invariant, meaning that they do not change if the viscosity of the shell is rescaled by a global factor.
They are thus independent of the viscosity of the shell if it is uniform.
\item 
They have the same thin shell limit as elastic isostasy (Eq.~(\ref{ShapeRatioTSI})) and as viscoelastic isostasy with constant load.
In other words, boundary conditions do not matter in the thin shell limit.
\item
At high harmonic degree, the shape ratio for a homogeneous shell tends to zero for pure top loading whereas it diverges to $-\infty$ for pure bottom loading (Eqs.~(\ref{SnVeSIAsymptoticTop})-(\ref{SnVeSIAsymptoticBot})).
The compensation factor does not diverge.
\end{enumerate}

%%%%%%%%%%%%%%%%%%%%%%%%%%%%%%%%%%%%%%%%%%%%%%%%%%%%%%%%%%%%%%%%%%%%%%%%%%%%%%%%%%%%%%%%%%%%%%%%%%%%%%%
%%%%%%%%%%%%%%%%%%%%%%%%%%%%%%%%%%%%%%%%%%%%%%%%%%%%%%%%%%%%%%%%%%%%%%%%%%%%%%%%%%%%%%%%%%%%%%%%%%%%%%%
\subsection{Variable load and shape}
\label{ViscoelasticVariableLoadShape}

Is isostatic balance still possible for loads that are neither constant nor proportional to time?
In the long time limit, isostatic balance can be approximately satisfied, or not at all, depending on whether the load variation is slower or faster than the time scale of viscoelastic relaxation.
As a first example, consider loading with polynomial time dependence $T(t)\sim t^q$ (see Table~\ref{TableLaplace}), of which the `constant shape' model is a special case ($q=1$).
In the long time limit, the deviatoric Love number behaves as
\begin{equation}
\hat h_j^{J}(t)
\sim  - \frac{t^{q-1}}{\Gamma(q)} \sum_p \frac{r^J_{jp}}{(s_p)^2} \, ,
\end{equation}
which is proportional to the `constant shape' value (Eq.~(\ref{hdevC})).
The resulting isostatic ratios are identical to those of the `constant shape' model.
Thus, the `constant shape' isostatic balance generally holds for loads with polynomial time dependence, which vary more slowly (in the long time limit) than viscoelastic exponential decay.

A more realistic model is a load with periodic time dependence $T(t)\sim\cos(\omega t)$ (see Table~\ref{TableLaplace}), which could result from the oscillation of the orbital eccentricity \citep{ojakangas1986,meyer2008}.
In the long time limit, the deviatoric Love number behaves as
\begin{equation}
\hat h_j^{J}(t) \sim \sum_p \frac{r^J_{jp}}{s_p} \,  \frac{\omega}{s_p^2+\omega^2} \left( \omega \, \cos(\omega t) + s_p \sin (\omega t) \right) .
\end{equation}
If the load varies much more slowly than the viscoelastic timescale ($\omega\ll |s_p|$), then the deviatoric Love number is well approximated in the long time limit by
\begin{equation}
\hat h_j^{J}(t) \sim  \omega \sin (\omega t)  \sum_p \frac{r^J_{jp}}{(s_p)^2} \, .
\end{equation}
In that case, the isostatic ratios are identical to those of the `constant shape' model.

%%%%%%%%%%%%%%%%%%%%%%%%%%%%%%%%%%%%%%%%%%%%%%%%%%%%%%%%%%%%%%%%%%%%%%%%%%%%%%%%%%%%%%%%%%%%%%%%%%%%%%%
%%%%%%%%%%%%%%%%%%%%%%%%%%%%%%%%%%%%%%%%%%%%%%%%%%%%%%%%%%%%%%%%%%%%%%%%%%%%%%%%%%%%%%%%%%%%%%%%%%%%%%%
%%%%%%%%%%%%%%%%%%%%%%%%%%%%%%%%%%%%%%%%%%%%%%%%%%%%%%%%%%%%%%%%%%%%%%%%%%%%%%%%%%%%%%%%%%%%%%%%%%%%%%%
\section{Viscous isostasy}
\label{SectionViscous}

%%%%%%%%%%%%%%%%%%%%%%%%%%%%%%%%%%%%%%%%%%%%%%%%%%%%%%%%%%%%%%%%%%%%%%%%%%%%%%%%%%%%%%%%%%%%%%%%%%%%%%%
%%%%%%%%%%%%%%%%%%%%%%%%%%%%%%%%%%%%%%%%%%%%%%%%%%%%%%%%%%%%%%%%%%%%%%%%%%%%%%%%%%%%%%%%%%%%%%%%%%%%%%%
\subsection{Slow viscous flow}
\label{SectionSlowViscousFlow}

The theory of slow viscous flow describes flows in which the fluid is incompressible, isothermal, and has a rheology with no memory of a past elastic response \citep{ribe2015}.
It has been widely applied -- with stationary boundary conditions -- to study plate dynamics, mantle convection, and geoid anomalies (\citet{hager1979,richards1984,ricard1984}; earlier work is reviewed by \citet{hager1978}).
Beside these stationary models, slow viscous flows have been used to compute time-dependent post-glacial rebound on Earth \citep{hager1979}  and the relaxation of impact basins on planetary bodies \citep{solomon1982}.
Two recent applications are directly relevant to the computation of isostatic ratios: the time-dependent approach used by \citet{ermakov2017} for Ceres, which is based on the propagator matrix method of \citet{hager1989}, and the stationary model proposed by \citet{cadek2019} for Enceladus, which they solve with their own semi-analytical (spectral) methods and fully numerical (finite-element) methods.

%%%%%%%%%%%%%%%%%%%%%%%%%%%%%%%%%%%%%%%%%%%%%%%%%%%%%%%%%%%%%%%%%%%%%%%%%%%%%%%%%%%%%%%%%%%%%%%%%%%%%%%
%%%%%%%%%%%%%%%%%%%%%%%%%%%%%%%%%%%%%%%%%%%%%%%%%%%%%%%%%%%%%%%%%%%%%%%%%%%%%%%%%%%%%%%%%%%%%%%%%%%%%%%
\subsection{Stokes-Rayleigh analogy}
\label{StokesRayleighAnalogy}

The fluid limit of elastic isostasy (Appendix~\ref{AppendixElasticIsostasy}) suggests that it does not differ that much from viscoelastic or viscous relaxation.
There is actually a precise mathematical correspondence, called the Stokes-Rayleigh analogy, between the theory of small incompressible elastic deformations and the theory of slow viscous flows \citep{ribe2015}.
The constitutive equations of the two theories are related by the following exchanges:
\begin{eqnarray*}
\mbox{displacements} & \leftrightarrow& \mbox{velocities} \, ,
\nonumber \\
\mbox{shear modulus} & \leftrightarrow& \mbox{viscosity} \, .
\nonumber
\end{eqnarray*}
This correspondence is not readily apparent when these theories are formulated in linearized form (as sets of 6 differential equations of first order in the propagator matrix approach), because the equations of motion involving viscous and elastic stresses are derived in Eulerian and Lagrangian descriptions, respectively \citep{wang1997}.

Beyond the governing equations, the Stokes-Rayleigh analogy also applies to the solutions to boundary value problems if the boundary conditions themselves are analogous \citep{ribe2018}.
In the case of interest here, the Stokes-Rayleigh analogy predicts for an incompressible body that elastic isostasy with zero surface (resp.\ bottom) deflection is equivalent to `constant shape' viscous isostasy with zero surface (resp.\ bottom) radial velocity.

%%%%%%%%%%%%%%%%%%%%%%%%%%%%%%%%%%%%%%%%%%%%%%%%%%%%%%%%%%%%%%%%%%%%%%%%%%%%%%%%%%%%%%%%%%%%%%%%%%%%%%%
%%%%%%%%%%%%%%%%%%%%%%%%%%%%%%%%%%%%%%%%%%%%%%%%%%%%%%%%%%%%%%%%%%%%%%%%%%%%%%%%%%%%%%%%%%%%%%%%%%%%%%%
\subsection{Propagator matrix approach}
\label{ViscousGeneralApproach}

The theory of slow viscous flow is very similar in structure to the gravitational-elastic equations used to compute Love numbers, with the important difference that displacement and strain variables are replaced by velocity and strain rate variables (this correspondence is discussed in Section~\ref{StokesRayleighAnalogy}).
In the formulation of \citet{hager1989}, there are 6 first-order differential equations for the radial and tangential velocities, the radial and radial-tangential stresses, and the gravitational potential and its derivative.
The problem is solved by propagating the variables at the bottom of the viscous shell to the surface, and by imposing the appropriate boundary conditions at both interfaces.
\citet{hager1989} do not give explicitly the propagator for the velocity and stress variables (suggesting to compute it with Sylvester's formula), while their propagator for the gravity variables is printed with typos.
For these reasons, and also to emphasize the similarity of the viscous formalism with the gravitational-elastic theory, I show in Appendix~\ref{AppendixViscousFlow} how to compute the propagator matrices as two-point products of the so-called fundamental matrices.
Moreover, the boundary conditions on (Eulerian) stress and gravity variables are not always explicit or appear with some typos in \citet{hager1989}; I give them in correct form in Appendix~\ref{AppendixViscousFlow}.
An implicit assumption of the viscous approach of \citet{hager1989} is that there is no perturbation due to a deformable solid core.

The procedure described in Appendix~\ref{ReducedPropagationSystem} yields two equations (or reduced propagation system) for the radial velocities $(u_1(R_s),u_1(R_o))$ and the radial displacements $(\delta{}r_s,\delta{}r_o)$.
If the shell is homogeneous, a fully analytical solution is feasible (see complementary software).
In that case, the reduced propagation system has nondimensional coefficients which do not depend on viscosity if displacement variables are normalized as
\begin{equation}
\delta{}\bar{r}_j=\delta{}r_j/\tau_s
\hspace{5mm} (j=s,o) \, ,
\end{equation}
where the time scale $\tau_s=\eta_s/(\rho_bg_sR_s)$ is defined in terms of the uniform viscosity $\eta_s$.

If the rheology of the shell depends on depth, the problem can be solved numerically with the propagator matrix method by modelling the shell as a superposition of homogeneous thin spherical layers.
The reduced propagation system is nondimensionalized as above, except that $\eta_s$ is now an arbitrarily chosen viscosity (for example the viscosity at the surface).
Moreover, the reduced propagation system is independent of the reference viscosity $\eta_0$ introduced for normalization purposes in the defining matrix ${\bf A}$ (Eq.~(\ref{ViscousMatA})).

%%%%%%%%%%%%%%%%%%%%%%%%%%%%%%%%%%%%%%%%%%%%%%%%%%%%%%%%%%%%%%%%%%%%%%%%%%%%%%%%%%%%%%%%%%%%%%%%%%%%%%%
%%%%%%%%%%%%%%%%%%%%%%%%%%%%%%%%%%%%%%%%%%%%%%%%%%%%%%%%%%%%%%%%%%%%%%%%%%%%%%%%%%%%%%%%%%%%%%%%%%%%%%%
\subsection{Constant load}
\label{SectionViscousConstantLoad}

\begin{figure}
   \centering
    \includegraphics[width=11.6cm]{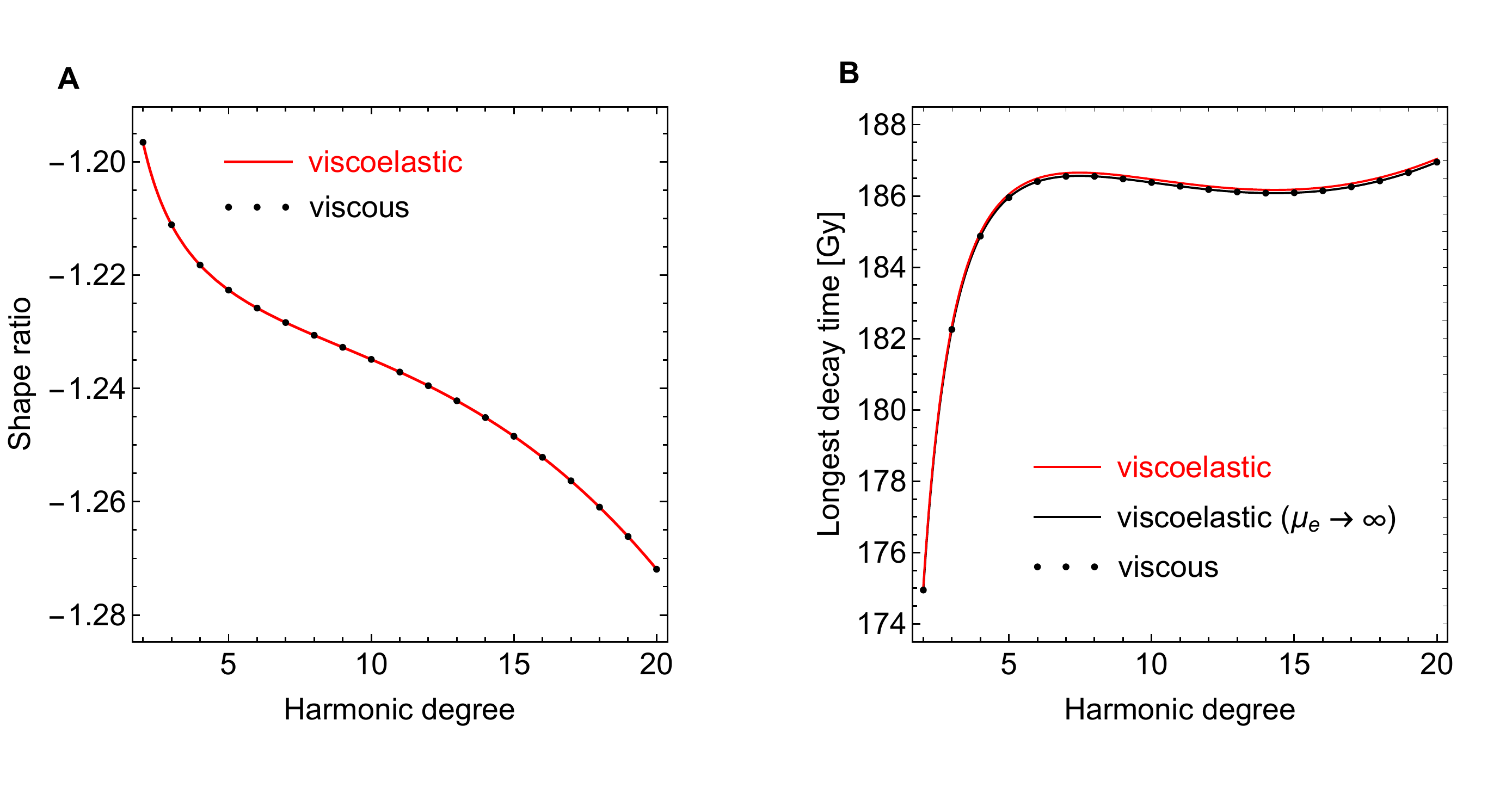}
   \caption[Viscoelastic/viscous isostasy with constant load: benchmarking]
   {Viscoelastic/viscous isostasy with constant load: benchmarking against the solution of \citet{ermakov2017}.
   The two panels show the shape ratio and the longest decay time as a function of harmonic degree.
   Parameters are the same as in Fig.~\ref{FigShapeGravTime}.
   Solid red curves show the predictions of viscoelastic isostasy; black dots are the predictions of viscous isostasy.
   Predictions for the shape ratio are analytically (and numerically) the same in the two approaches.
   Viscoelastic decay times are slightly affected by the elastic shear modulus $\mu_{\rm e}$, but become identical to viscous decay times in the limit of large $\mu_{\rm e}$ (only the longest decay time is shown).}
   \label{FigViscoCeres}
\end{figure}

Similarly to viscoelastic isostasy, viscous isostasy can be solved for different loading histories.
If the load is constant, the positions of the boundaries depend on time, as done for glacial unloading by \citet{hager1979} and for the isostatic shape ratio by \citet{ermakov2017}.
The condition of zero radial velocity imposed by \citet{hager1989} is thus not appropriate here.
Instead, the radial velocity of the flow is set equal to the time-derivative of the radial displacement:
\begin{equation}
u_1(R_s) =\frac{\partial}{\partial t}  ( \tau_s \, \delta{}\bar{r}_s )
\hspace{5mm} \mbox{and}  \hspace{5mm}
u_1(R_o) = \frac{\partial}{\partial t}  ( \tau_s \,  \delta{}\bar{r}_o ) \, .
\end{equation}
With these conditions, the reduced propagation system can be written as a system of two coupled homogeneous differential equations of the first order for the time-dependent functions $\delta{}r_s(t)$ and $\delta{}r_o(t)$:
\begin{equation}
\tau_s \, \frac{\partial}{\partial t}
\left(
\begin{array}{c}
\delta{}\bar{r}_s(t) \\
\delta{}\bar{r}_o(t)
\end{array}
\right)
=
\mathbf{M} \cdot
\left(
\begin{array}{c}
\delta{}\bar{r}_s(t) \\
\delta{}\bar{r}_o(t)
\end{array}
\right) ,
\label{eigeneq}
\end{equation}
where $\mathbf{M}$ is a $2\times2$ matrix depending on nondimensional parameters (it does not depend on viscosity if the shell is homogeneous).
For example, the parameters for a homogeneous shell can be chosen as $(x,\xi_{s1},\xi_{o1},n)$.
Eq.~(\ref{eigeneq}) has two independent solutions given by
\begin{equation}
\left(
\begin{array}{c}
\delta{}\bar{r}_s(t) \\
\delta{}\bar{r}_o(t)
\end{array}
\right)
=
\left(
\begin{array}{c}
w_{s\pm} \\
w_{o\pm}
\end{array}
\right)
\exp(\lambda_\pm{}t/\tau_s) \, ,
\end{equation}
where $\lambda_\pm$ and $\mathbf{w_\pm}=(w_{s\pm} \, w_{o\pm})$ are the eigenvalues and eigenvectors of $\mathbf{M}$, with $\lambda_-<\lambda_+<0$.
The short ($-$) and long ($+$) decay times are given by
\begin{equation}
\tau_\pm = -\tau_s/\lambda_\pm \, .
\label{ViscousDecayTimes}
\end{equation}
At large times ($t\gg\tau_s/|\lambda_-|$), the solution with the most negative eigenvalue $\lambda_-$ becomes negligible and the shape ratio tends to a constant:
\begin{equation}
S_n^{\rm VLI} = \lim_{t\rightarrow\infty} \frac{\delta{}\bar{r}_o(t)}{\delta{}\bar{r}_s(t)} = \frac{w_{o+}}{w_{s+}} \, ,
\end{equation}
where `VLI' stands for \textbf{V}iscous (constant) \textbf{L}oad \textbf{I}sostasy.
If the shell is homogeneous, this method yields analytical formulas for the decay times and the shape ratio, which depend on the nondimensional parameters $(x,\xi_{s1},\xi_{so},n)$, but not on viscosity.
Assuming an incompressible body, I show in the complementary software that these formulas are identical to the `constant load' viscoelastic isostatic ratios of Section~\ref{ViscoelasticConstantLoad}.
They also agree perfectly with the numerical results of \citet{ermakov2017} (see Fig.~\ref{FigViscoCeres}).

%%%%%%%%%%%%%%%%%%%%%%%%%%%%%%%%%%%%%%%%%%%%%%%%%%%%%%%%%%%%%%%%%%%%%%%%%%%%%%%%%%%%%%%%%%%%%%%%%%%%%%%
%%%%%%%%%%%%%%%%%%%%%%%%%%%%%%%%%%%%%%%%%%%%%%%%%%%%%%%%%%%%%%%%%%%%%%%%%%%%%%%%%%%%%%%%%%%%%%%%%%%%%%%
\subsection{Constant shape}
\label{SectionViscousConstantShape}

For Enceladus, \citet{cadek2019} assume that the shape of the shell-ocean boundary is maintained by melting or freezing due to heat flow from below the shell.
This shape  cannot be identified with the bottom boundary of the viscous layer, because there is a continuous transfer of material between the shell and the ocean.
One can impose, however, that the surface shape does not change with time (since there is no transfer of material there) or, equivalently, that the radial velocity of the surface vanishes:
\begin{equation}
 u_1(R_s)=0 \, .
\end{equation}
In that model, the radial velocity at the bottom corresponds to the radial velocity of the ice flow, which does not vanish but is compensated by the formation or melting of ice.
Contrary to the case of constant load, the velocity of the flow at the bottom boundary is not directly related to the radial displacement.
The reduced propagation system of Appendix~\ref{ReducedPropagationSystem} can be written as two coupled inhomogeneous equations in the variables $(u_1(R_o),\delta{}\bar{r}_o)$ with an inhomogeneous term proportional to $\delta{}\bar{r}_s$ (it can be set equal to 1).
Once this system has been solved, the isostatic shape ratio is given by
\begin{equation}
S_n^{\rm VSI} = \frac{\delta{}\bar{r}_o}{\delta{}\bar{r}_s} \, ,
\label{SnVSI}
\end{equation}
where `VSI' stands for \textbf{V}iscous (constant) \textbf{S}hape \textbf{I}sostasy.
If the shell is homogeneous, the shape ratio does not depend on the uniform viscosity of the shell.
If the shell is stratified into thin layers of homogeneous rheology, the shape ratio depends on the ratios of the viscosities of the different layers, but is invariant under an overall rescaling of the shell viscosity.
This property of \textit{$\eta$-invariance} mirrors the property of $\mu$-invariance for elastic isostasy.

For an incompressible body, the VSI shape ratio given by Eq.~(\ref{SnVSI}) is analytically equivalent to the VeSI shape ratio with purely bottom loading (see complementary software).
Therefore, it is equivalent to the shape ratio of elastic isostasy with zero deflection at the surface, exactly as predicted by the Stokes-Rayleigh analogy (Section~\ref{StokesRayleighAnalogy}).
Mathematically, it is possible to define viscous isostasy with zero velocity at the shell-ocean boundary, although the physical interpretation is not obvious.
The linear combination of the two extreme boundary conditions results in a 1-parameter isostatic family which is equivalent to the 1-parameter family of zero-deflection elastic isostasy.

\begin{figure}
\centering
    \includegraphics[width=11.6cm]{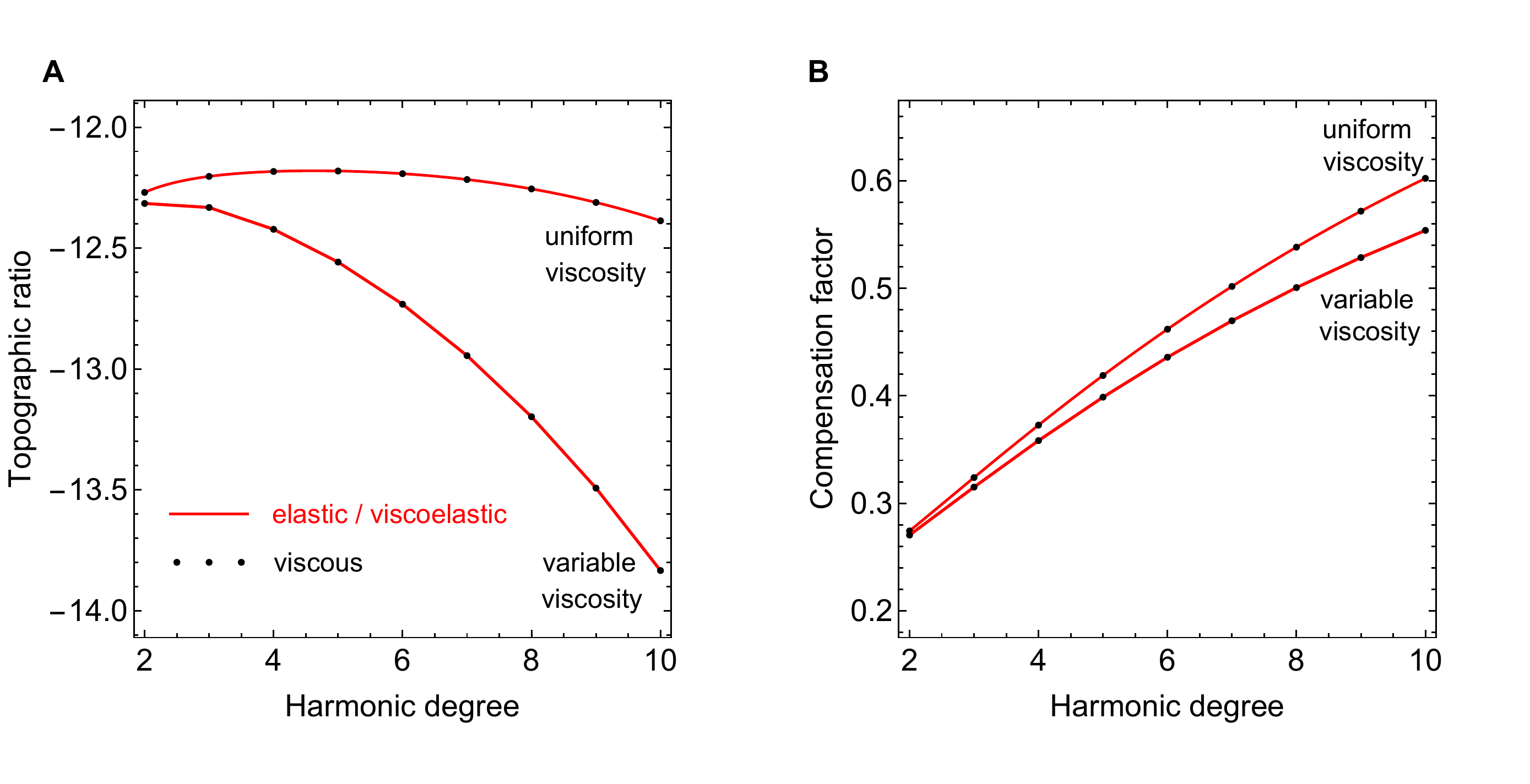}
   \caption[Elastic/viscoelastic/viscous isostasy with constant shape: benchmarking]
   {Elastic/viscoelastic/viscous isostasy with constant shape: benchmarking against the solution of \citet{cadek2019}.
   The two panels show the topographic ratio and the compensation factor as a function of harmonic degree.
   The 3-layer model of Enceladus is specified by $\rho_s=926\rm\,kg/m^3$, $\rho_o=1010\rm\,kg/m^3$, $d=20\rm\,km$, $R_s=252.1\rm\,km$, and a non-deformable core with density $2366\rm\,kg/m^3$ and radius $194\rm\,km$.
   Viscoelastic isostasy is computed from elastic isostasy with no surface deflection ($\alpha=0$) given by Eq.~(\ref{TopoRatioZDI}) if viscosity is uniform; the variable viscosity case is solved numerically with the elastic propagator matrix method \citep{sabadini2016}.
   Viscous isostasy is solved with the viscous propagator matrix method (Section~\ref{SectionViscousConstantShape}).
   For clarity, the elastic/viscoelastic solution is shown as a continuous curve.}
   \label{FigBenchmarkCadek}
\end{figure}

Finally, the results obtained with the propagator matrix method described here agree perfectly with the results of \citet{cadek2019} (more precisely those obtained with the spectral method), whether the shell has a constant or variable viscosity (Fig.~\ref{FigBenchmarkCadek}).
If the shell has a depth-dependent viscosity, the equivalent elastic model consists in a shell having a shear modulus that depends on depth in the same way as the viscosity in the viscous model (note that the overall scale of the shear modulus does not matter because of $\mu$-invariance).
Fig.~\ref{FigBenchmarkCadek} shows that this procedure yields the same results as the viscous propagator method.

%%%%%%%%%%%%%%%%%%%%%%%%%%%%%%%%%%%%%%%%%%%%%%%%%%%%%%%%%%%%%%%%%%%%%%%%%%%%%%%%%%%%%%%%%%%%%%%%%%%%%%%
%%%%%%%%%%%%%%%%%%%%%%%%%%%%%%%%%%%%%%%%%%%%%%%%%%%%%%%%%%%%%%%%%%%%%%%%%%%%%%%%%%%%%%%%%%%%%%%%%%%%%%%
%%%%%%%%%%%%%%%%%%%%%%%%%%%%%%%%%%%%%%%%%%%%%%%%%%%%%%%%%%%%%%%%%%%%%%%%%%%%%%%%%%%%%%%%%%%%%%%%%%%%%%%
\section{Application to Enceladus and Europa}
\label{SectionDiscussion}

%%%%%%%%%%%%%%%%%%%%%%%%%%%%%%%%%%%%%%%%%%%%%%%%%%%%%%%%%%%%%%%%%%%%%%%%%%%%%%%%%%%%%%%%%%%%%%%%%%%%%%%
%%%%%%%%%%%%%%%%%%%%%%%%%%%%%%%%%%%%%%%%%%%%%%%%%%%%%%%%%%%%%%%%%%%%%%%%%%%%%%%%%%%%%%%%%%%%%%%%%%%%%%%
\subsection{Interior models and computational methods}
\label{SectionExampleCases}

As in Paper~I, I consider Enceladus and Europa as example cases of isostasy.
Both bodies harbour an internal ocean but differ significantly in their shell thickness: Enceladus has a thick shell ($d_s\sim8$ to 12\% of $R_s$), whereas Europa has a thin shell ($d_s\sim1$ to 5\% of $R_s$).
The interior model for Enceladus is specified by $\rho_s=920\rm\,kg/m^3$; $\rho_o=1020\rm\,kg/m^3$; $\rho_b=1610\rm\,kg/m^3$; $d=23\rm\,km$; $R_s=252.1\rm\,km$ \citep{beuthe2016}.
For Europa, I adopt the same shell density, ocean density, and shell thickness as for Enceladus; the bulk density and the surface radius are $\rho_b=3013\rm\,kg/m^3$ and $R_s=1560.8\rm\,km$.
The core is infinitely rigid in both cases.

I will first assume that the shell is of uniform viscosity, in which case viscosity drops out of the isostatic ratios and analytical formulas are available (see complementary software).
This model corresponds closely to what is traditionally viewed as Airy isostasy.
I will then examine the more realistic case of a shell with depth-dependent rheology (the `constant shape' case was previously studied by \citet{cadek2019}; see Fig.~\ref{FigBenchmarkCadek}).
In that case, isostatic ratios are affected by viscosity variations but do not depend on the overall magnitude of viscosity.
This model must be solved numerically. 
Stationary viscoelastic isostasy is equivalent to ZDI elastic isostasy and can thus be solved with the propagator matrix method for elastic Love numbers \citep{sabadini2016} or, alternatively, with the viscous propagator method of Section~\ref{SectionViscousConstantShape}.
Time-dependent isostasy can be solved either by computing Love number residues \citep{peltier1985,jaraorue2011,sabadini2016} or with the viscous propagator method of Section~\ref{SectionViscousConstantLoad}.

%%%%%%%%%%%%%%%%%%%%%%%%%%%%%%%%%%%%%%%%%%%%%%%%%%%%%%%%%%%%%%%%%%%%%%%%%%%%%%%%%%%%%%%%%%%%%%%%%%%%%%%
%%%%%%%%%%%%%%%%%%%%%%%%%%%%%%%%%%%%%%%%%%%%%%%%%%%%%%%%%%%%%%%%%%%%%%%%%%%%%%%%%%%%%%%%%%%%%%%%%%%%%%%
\subsection{Uniform viscosity}

\begin{figure}
\centering
    \includegraphics[width=12cm]{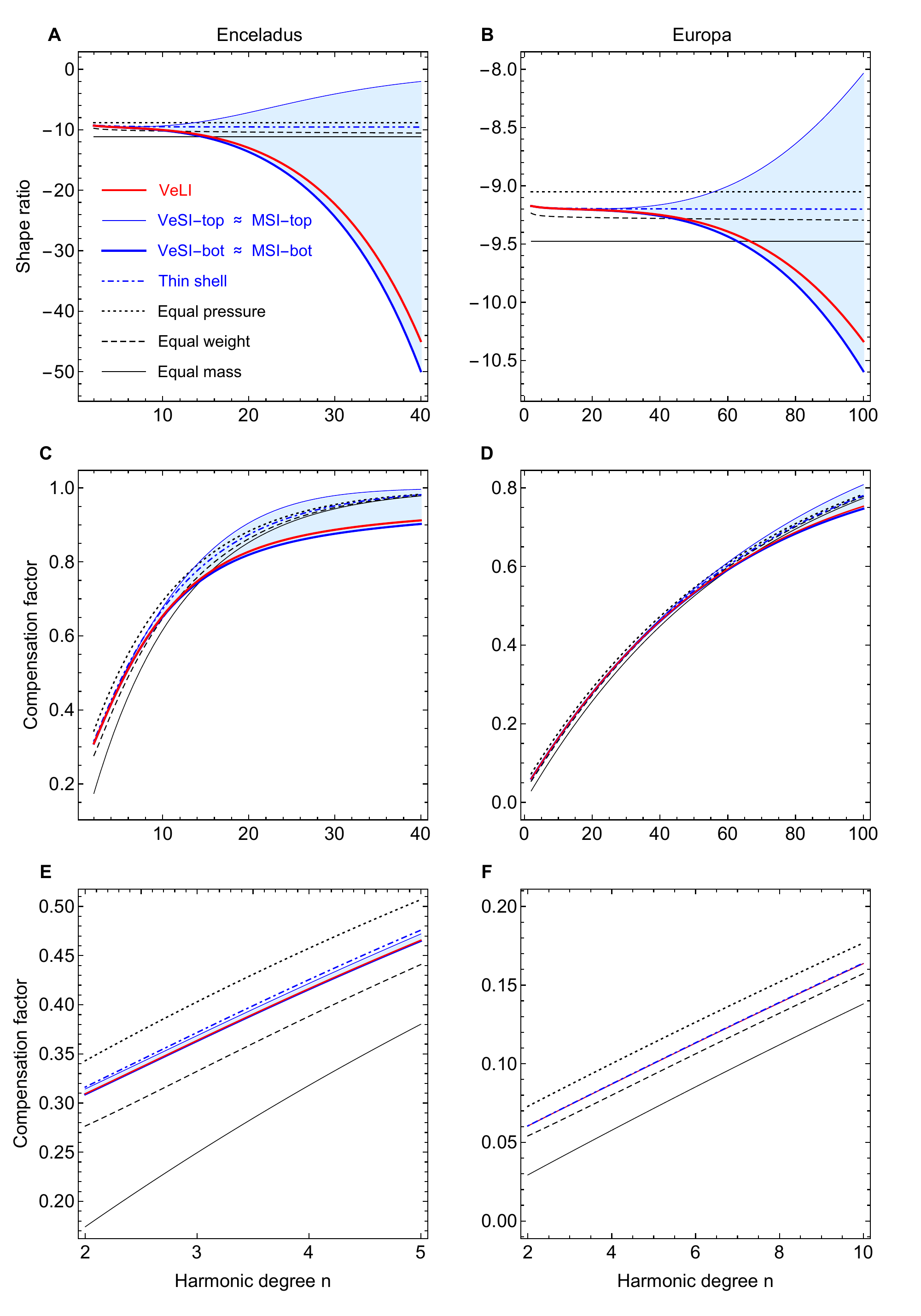}
   \caption[Viscoelastic isostasy for a homogeneous shell: application to Enceladus and Europa]{
   Viscoelastic isostasy for a homogeneous shell: shape ratio (first row) and compensation factor (second and third rows), as a function of harmonic degree, for Enceladus (left) and Europa (right).
   Lower panels zoom on the lower harmonic degrees of the middle panels.
   Red curves show `constant load' viscoelastic isostasy (VeLI).
   Blue curves show `constant shape'  viscoelastic isostasy (VeSI), with thick and thin curves corresponding to bottom (= bot) and top loading, respectively.
   The shaded area represents combined top and bottom loading.
  The connection to minimum stress isostasy (MSI) is discussed in the text.
  Various models of classical isostasy are shown as thin black curves: equal pressure (dotted), equal weight (dashed), equal mass (solid).
   `Improved' thin shell isostasy (see after Eq.~(\ref{ShapeRatioTSI})) is shown as a dash-dotted blue curve.
   Interior models are specified in Section~\ref{SectionExampleCases}.
 }
   \label{FigCompaViscoUniform}
\end{figure}

\begin{figure}
\centering
    \includegraphics[width=6cm]{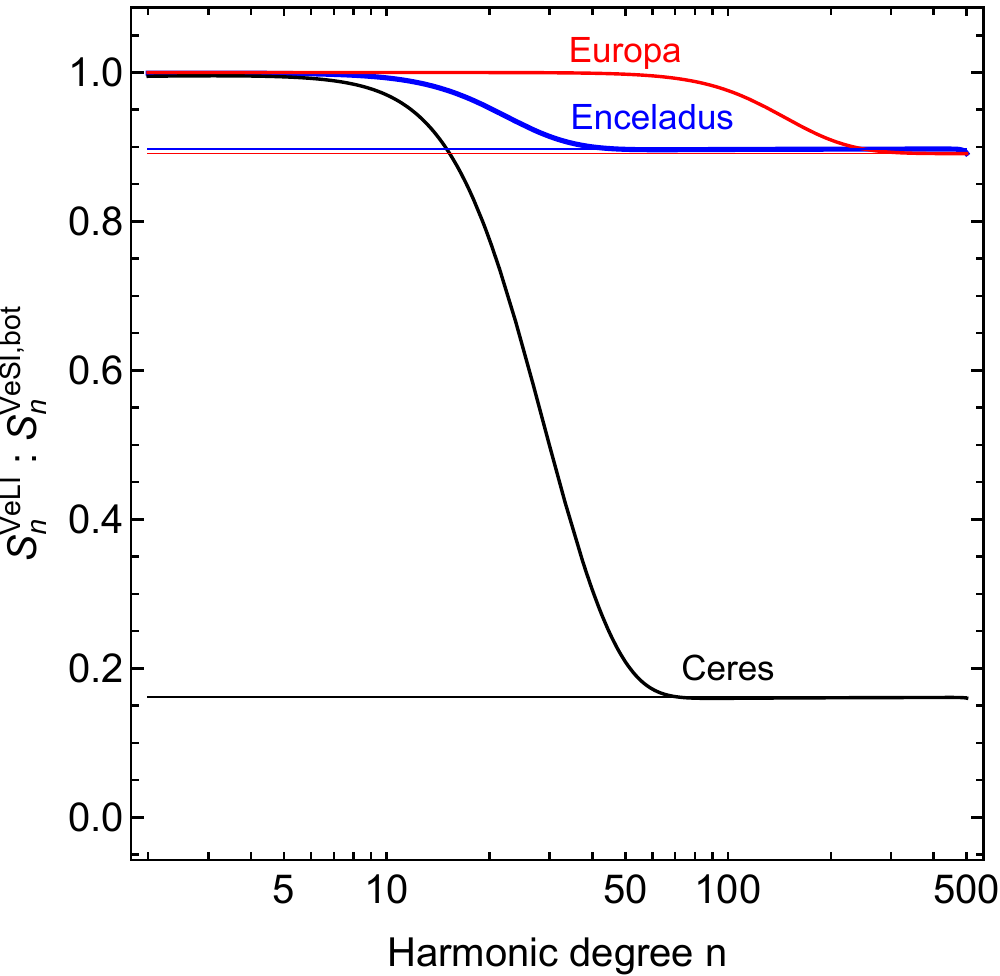}
   \caption[Asymptotic behaviour of the VeLI shape ratio for a homogeneous shell]
   {Asymptotic behaviour of the VeLI shape ratio for a homogeneous shell.
At high harmonic degree, the VeLI shape ratio tends to a constant fraction of the VeSI shape ratio.
The constant fraction (shown as thin horizontal lines) is equal to $1-\gamma_o{}x\Delta\xi_1/\xi_{s1}$ (Eq.~(\ref{SnVeLIAsymptoticRatio1})). }
   \label{FigAsymptotic}
\end{figure}

Fig.~\ref{FigCompaViscoUniform} shows the viscoelastic shape ratio and compensation factor for Enceladus and Europa as a function of harmonic degree if their shell is of uniform viscosity.
Predictions of viscoelastic isostasy with `constant load' (VeLI) or `constant shape' (VeSI) are indistinguishable at low harmonic degree and are well approximated in that range by thin shell isostasy (using the `improved' version of thin shell isostasy discussed after Eq.~(\ref{ShapeRatioTSI})).
As harmonic degree increases, the shape ratio diverges for VeLI as well as for VeSI-bottom (that is VeSI with bottom loading, which is the physically relevant case).
At high harmonic degree, VeLI remains close to VeSI-bottom if the shell-ocean density contrast is small, whereas VeLI diverges more slowly if the density contrast is large (as in the Ceres model of Fig.~\ref{FigViscoCeres}).
This statement can be made more precise by studying the asymptotic behaviour of the shape ratio at high harmonic degree (see Appendix~\ref{AppendixAsymptotic}).
For a 3-layer incompressible body with homogeneous layers, $S_n^{\rm VeLI}$ typically tends to a constant fraction of $S_n^{\rm VeSI, bot}$ (Fig.~\ref{FigAsymptotic}).
As in elastic isostasy, the divergence of the VeSI-bottom shape ratio at high harmonic degree can be explained by ``Jeffreys' theorem" \citep{melosh2011}: loads are supported by stresses over a radial distance of about $R_s/n\approx\lambda/2\pi$ where $\lambda$ is the load wavelength (see Fig.~9 of Paper~I).
Since a bottom load of small extent is supported by stresses close to the bottom of the shell, it does not induce stresses close to the surface and the surface shape gets smaller, compared to the bottom shape, as the wavelength gets shorter.

The divergence of the VeLI and VeSI-bottom shape ratios is much faster if the shell is thicker (compare left and right panels of Fig.~\ref{FigCompaViscoUniform}).
Simply put, the predictions for Europa look like the predictions for Enceladus at a smaller harmonic degree.
As expected, loads are more compensated ($F_n$ closer to zero) at a given harmonic degree if the shell is thin.

Given that the viscosity is uniform, the VeSI/VSI-bottom model is equivalent to elastic isostasy with zero surface deflection in a shell with uniform shear modulus, which is in turn nearly identical to elastic isostasy based on minimum stress with the bottom shape kept constant during minimization (thick blue curves in Fig.~\ref{FigCompaViscoUniform}).
It is more natural, however, to keep the surface shape constant during minimization since it is more readily observable than the bottom boundary (thin blue curves in Fig.~\ref{FigCompaViscoUniform}).
In that case, the predictions of the VeSI/VSI-bottom model differ from those of minimum stress isostasy.
Nevertheless, the difference is only significant at short wavelengths where little Airy compensation occurs.

Predictions of classical isostasy are significantly off at long wavelengths, especially for the compensation factor (Fig.~\ref{FigCompaViscoUniform}, panels E and F).
In classical isostasy, the shape ratio varies little with the harmonic degree, which is a common feature of isostasy based on local compensation (that is, vertical columns move independently).

%%%%%%%%%%%%%%%%%%%%%%%%%%%%%%%%%%%%%%%%%%%%%%%%%%%%%%%%%%%%%%%%%%%%%%%%%%%%%%%%%%%%%%%%%%%%%%%%%%%%%%%
%%%%%%%%%%%%%%%%%%%%%%%%%%%%%%%%%%%%%%%%%%%%%%%%%%%%%%%%%%%%%%%%%%%%%%%%%%%%%%%%%%%%%%%%%%%%%%%%%%%%%%%
\subsection{Depth-dependent rheology}
\label{SectionRheology}

At first sight, icy satellites with subsurface oceans provide a perfect setting for Airy isostasy, because the fluid layer below the compensation depth is not just an idealization but a fact.
Complications however arise because such an icy shell cannot be approximated as homogeneous: the bottom of the shell is close to the melting temperature of ice and has thus a soft rheology, whereas the top of the shell is extremely cold and thus nearly rigid.
Following \citet{beuthe2018,beuthe2019}, I assume that the shell is in a conductive state with conductivity inversely proportional to temperature.
For a shell of thickness $d$ without internal heat production, the temperature $T$ at depth $z$ from the surface is given by $T/T_s=(T_m/T_s)^{z/d}$ (spherical corrections are neglected), where $T_m$ is the melting temperature ($273\,$K) at the base of the shell and $T_s$ is the mean surface equilibrium temperature ($59\,$K for Enceladus and $100\,$K for Europa).
The viscosity of ice is related to temperature by an Arrhenius relation: $\eta=\eta_m\exp(E_a(1/T-1/T_m))/R_g$, where $\eta_m=10^{14}\rm\,Pa.s$ is the viscosity at the melting temperature, $E_a=59.4\rm\,kJ\,mol^{-1}$ is the activation energy for diffusion creep, and $R_g$ is the gas constant.
Close to the melting temperature, this rudimentary model is only a rough approximation of a much more complex picture \citep{goldsby2001}.
At low temperatures, laboratory constraints are lacking \citep{durham2001} and it is customary (for numerical convenience) to impose a viscosity cutoff in the range $10^{20}$ to $10^{30}\rm\,Pa.s$.

\begin{figure}
\centering
    \includegraphics[width=12cm]{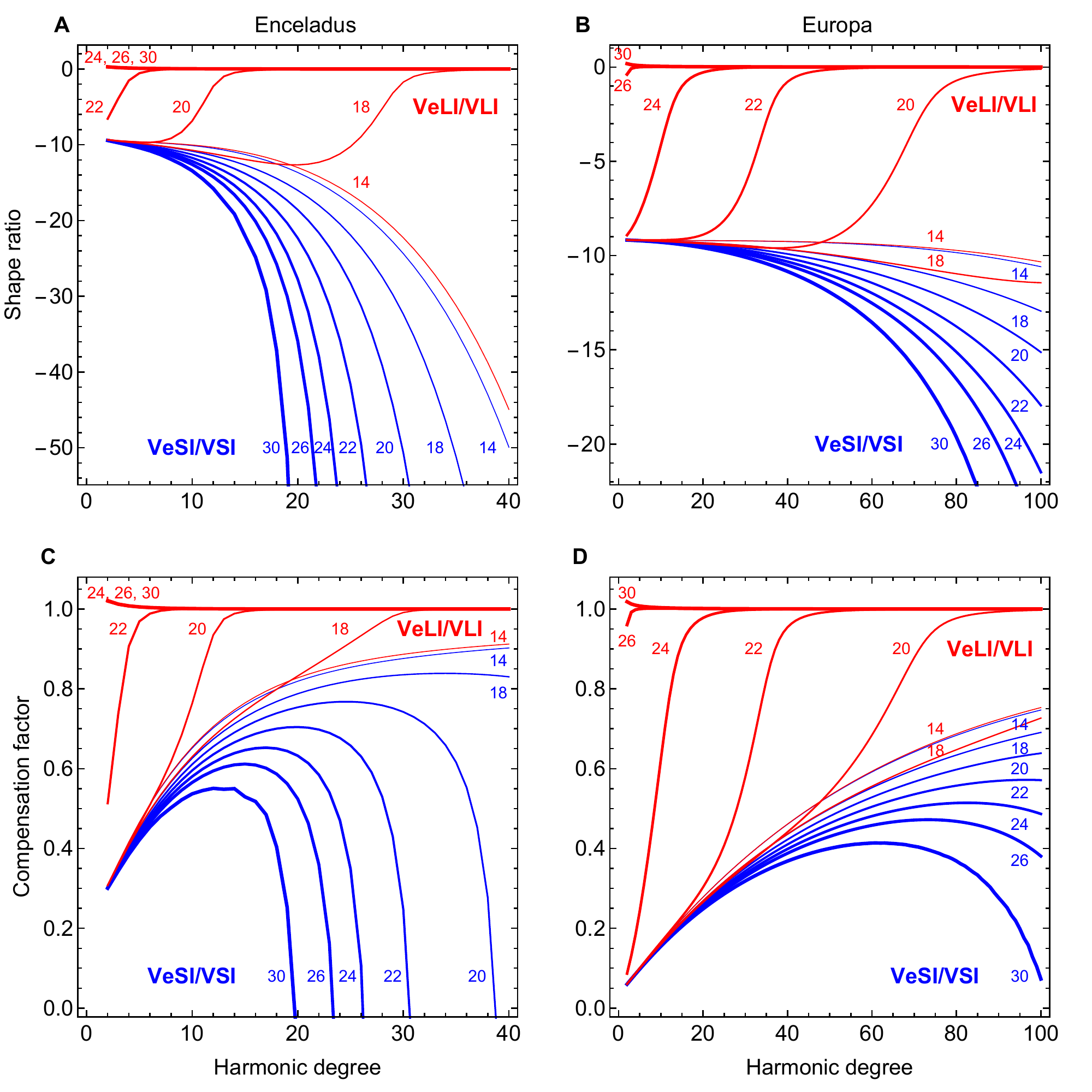}
   \caption[Viscoelastic isostasy for a shell with depth-dependent rheology: application to Enceladus and Europa]
   {Viscoelastic isostasy for a shell with depth-dependent rheology: shape ratio (first row) and compensation factor (second row), as a function of harmonic degree, for Enceladus (left) and Europa (right).
   Red curves show isostasy with constant load (VeLI or VLI) whereas blue curves show isostasy with constant bottom shape (VeSI or VSI).
   The bottom viscosity is $10^{14}\rm\,Pa.s$.
   The thickness of the curve increases with the cutoff on the maximum viscosity $\eta_{\rm cut}$ (labeled by $\log_{10}(\eta_{\rm cut})$).
   The label `14' corresponds to a homogeneous shell as on Fig.~\ref{FigCompaViscoUniform}.
   Interior models of Enceladus and Europa are the same as in Fig.~\ref{FigCompaViscoUniform}; the rheology is specified in Section~\ref{SectionRheology}.}
   \label{FigCompaViscoVariable}
\end{figure}

Fig.~\ref{FigCompaViscoVariable} shows the VeLI and VeSI-bottom shape ratio and compensation factor for Enceladus and Europa as a function of harmonic degree.
If the load is constant, isostatic ratios are radically affected by the depth-dependent rheology: Airy compensation becomes negligible at all wavelengths if the viscosity cutoff is high enough ($10^{24}-10^{26}\rm\,Pa.s$).
If the bottom shape is constant, the shape ratio diverges faster with harmonic degree than for a shell of uniform viscosity, confirming what \citet{cadek2019} found for the topographic ratio (see Fig.~\ref{FigBenchmarkCadek}).
The shape ratio even becomes singular at a finite value of the harmonic degree.
Beyond that threshold, surface topography of positive amplitude cannot be supported by a stationary buoyant root at the bottom of the shell: Airy isostasy breaks down.
Nevertheless, the predictions of VeSI-bottom remain close to the uniform viscosity model at low harmonic degrees where the mechanism of heat transfer maintaining a stationary shape is the most plausible (less than 20\% deviation if $n\leq10$ for Enceladus and $n\leq50$ for Europa).

%%%%%%%%%%%%%%%%%%%%%%%%%%%%%%%%%%%%%%%%%%%%%%%%%%%%%%%%%%%%%%%%%%%%%%%%%%%%%%%%%%%%%%%%%%%%%%%%%%%%%%%
%%%%%%%%%%%%%%%%%%%%%%%%%%%%%%%%%%%%%%%%%%%%%%%%%%%%%%%%%%%%%%%%%%%%%%%%%%%%%%%%%%%%%%%%%%%%%%%%%%%%%%%
%%%%%%%%%%%%%%%%%%%%%%%%%%%%%%%%%%%%%%%%%%%%%%%%%%%%%%%%%%%%%%%%%%%%%%%%%%%%%%%%%%%%%%%%%%%%%%%%%%%%%%%
\section{Conclusions}

The application of Airy isostasy to a planet or moon at the global scale requires going beyond the classical picture of vertical columns floating in a fluid.
For this purpose, two kinds of new isostatic models have been developed: they are based, on the one hand, on static elastic equilibrium minimizing the crustal stress and, on the other, on dynamic equilibrium realized through time-dependent or stationary viscous flow.
After studying elastic models in Paper~I, I have examined in this sequel dynamic models of isostasy and how they can be related to elastic isostasy.
Among the various possible loading histories, two dynamic models stand out by their close connection to realistic scenarios.
In the first one, a constant load is applied at the surface.
This model describes topography that was emplaced in the past and has relaxed to a state of Airy balance, for example a crater or a dormant volcano.
In the second model, a time-dependent load is applied at the bottom of the shell so that the shape of the shell boundaries remains constant.
This stationary model corresponds to a dynamic equilibrium between viscous flow and melting/freezing at the bottom of the shell due to heat transfer between shell and ocean.
If heat is generated by tidal heating, it is likely that such an equilibrium mainly occurs at the longest wavelengths (harmonic degrees 2 and 4), although turbulence could result in melting/freezing variations of shorter wavelength if the ocean transports heat from the core to the shell.
The stationary model can be extended to scenarios where both load and shape vary slowly, as is the case if the thermal-orbital evolution of a satellite in a resonance falls into an oscillating pattern.

\begin{table}[h]
\hspace*{-3cm}
\ra{1.3}
\footnotesize
\caption[Equivalences between endmembers of elastic and dynamic approaches]{Equivalences between three endmembers of elastic and dynamic approaches to isostasy.
Equivalent models are grouped in three rows.}
\vspace{1.5mm}
\begin{tabular}{@{}lllll@{}}
\hline
\vspace{0.3mm}
& Elastic isostasy &  Elastic isostasy & Viscoelastic isostasy & Viscous isostasy\\
\hline
1 & zero deflection  & minimum stress &  constant shape & stationary (constant shape) \\
& no surface deflection & bottom shape nearly fixed &  increasing bottom load & no mass exchange at surface\\
& $\alpha=0$ & $\beta=\alpha_0<<0^{\,\,(*)}$ &  $\zeta_n^V=\pm\infty$ & $u_1(R_s)=0$ \\
\hline
2 & zero deflection  & minimum stress &  constant shape & stationary (constant shape) \\
& no bottom deflection & surface shape nearly fixed &  increasing surface load & no mass exchange at bottom \\
& $\alpha=\pm\infty$ & $\beta=\alpha_\infty\approx0^{\,\,(*)}$ &  $\zeta_n^V=0$ & $u_1(R_o)=0$ \\
\hline
3 & no elastic equivalent & no elastic equivalent & constant load  & time-dependent (constant load) \\
& -- & -- & time-dependent shape & no mass exchange at all \\
& -- & -- & loading ratio is irrelevant & unique boundary conditions\\
\hline
\multicolumn{5}{l}{\scriptsize ${}^{(*)}$ duality based on the homogeneous shell assumption (Eq.~(\ref{alphaMSI})).}
\end{tabular}
\label{TableEquiv}
\end{table}%

The first key result of this paper is to establish equivalences, wherever they exist, between elastic, viscoelastic, and viscous approaches to isostasy.
Fig.~\ref{FigIsoRelations} and Table~\ref{TableEquiv} summarize the results.
First, viscoelastic and viscous approaches do not differ at all.
Second, stationary viscoelastic/viscous isostasy is mathematically equivalent to zero-deflection elastic isostasy.
Third, for each model of stationary viscoelastic/viscous isostasy (assuming a homogeneous shell), one can find an equivalent model of minimum stress elastic isostasy, although the boundary conditions (given by $\zeta_n^V$ and $\beta$) are connected in a non-trivial way by the $\alpha-\beta$ duality discussed in Paper~I.

Thanks to these equivalences, stationary dynamic isostasy can be solved indifferently with elastic, viscoelastic, or viscous methods, depending on the user's preferences.
For time-dependent dynamic isostasy, searching for all the normal modes associated with Love numbers is probably not the most efficient method, since only the mode with the longest decay time is relevant to isostasy.
On the other hand, such codes have been developed to a high degree of sophistication for post-glacial rebound studies.
In the end, the choice of the method is more a question of software availability and personal experience.
For the viscous approach, I gave a new formulation of the propagation matrix in terms of the fundamental matrix, similar to what is done when computing elastic or viscoelastic Love numbers of an incompressible body stratified in homogeneous layers.
The viscous propagator matrices and the boundary conditions are fully implemented in the complementary software.

The existence of these equivalences teaches us two things.
On the one hand, the freedom to define a new isostatic model is limited: isostatic approaches based on different physical principles (elastic equilibrium or viscoelastic relaxation) yield exactly the same isostatic ratios if boundary conditions correspond (Table~\ref{TableEquiv}), and share the same thin shell limit whatever the boundary conditions (contrary to classical isostasy).
On the other, it shows that boundary conditions matter.
Isostatic ratios usually differ if boundary conditions are applied at the surface or at the bottom of the shell, with the exception of dynamic isostasy with constant load.
This difference, which is most notable in the shape ratio (Fig.~\ref{FigCompaViscoUniform}, panels A and B), increases with harmonic degree and with shell thickness, and depends on the assumption of isotropic elasticity (the difference is negligible if compensation is local).
It can be explained by the migration of supporting stresses closer to the applied load (``Jeffreys' theorem").
Predictions for the compensation factor are not very sensitive to the choice of boundary conditions if the shell is homogeneous (Fig.~\ref{FigCompaViscoUniform}, panels C to F).
If the rheology depends strongly on depth, one must think first about choosing the correct boundary conditions applicable to the problem and the harmonic degree range in which they are valid (Fig.~\ref{FigCompaViscoVariable}).

The second key result of this paper is to provide analytical formulas for viscoelastic/viscous isostasy, with either constant load or constant shape, for an incompressible 3-layer model with homogeneous layers (see code in \citet{beuthe2020z}).
It is thus not justified anymore (at least at long wavelengths) to work with classical isostasy formulas just because they are simpler.
If the shell is not of uniform viscosity, the analytical formulas can be used to benchmark the numerical codes used to compute isostatic ratios.

\small
\section*{Acknowledgments}
I thank Anton Ermakov and Ondrej Cadek for stimulating discussions which inspired this paper.
I also thank Isamu Matsuyama and Bert Vermeersen for constructive comments about the manuscript.
All data used in the paper are publicly available.
Mathematica and Fortran codes are available on https://zenodo.org (see \citet{beuthe2020z} in the reference list).
This work is financially supported by the Belgian PRODEX program managed by the European Space Agency in collaboration with the Belgian Federal Science Policy Office.

\normalsize
\begin{appendices}

%%%%%%%%%%%%%%%%%%%%%%%%%%%%%%%%%%%%%%%%%%%%%%%%%%%%%%%%%%%%%%%%%%%%%%%%%%%%%%%%%%%%%%%%%%%%%%%%%%%%%%%
%%%%%%%%%%%%%%%%%%%%%%%%%%%%%%%%%%%%%%%%%%%%%%%%%%%%%%%%%%%%%%%%%%%%%%%%%%%%%%%%%%%%%%%%%%%%%%%%%%%%%%%
%%%%%%%%%%%%%%%%%%%%%%%%%%%%%%%%%%%%%%%%%%%%%%%%%%%%%%%%%%%%%%%%%%%%%%%%%%%%%%%%%%%%%%%%%%%%%%%%%%%%%%%
\section{Elastic isostasy}
\label{AppendixElasticIsostasy}
\renewcommand{\theequation}{A.\arabic{equation}} % redefine the command that creates the equation no.
\setcounter{equation}{0}  % reset counter 

This section lists a few key results of Paper~I that are specific to elastic isostasy.
Elastic isostatic ratios are invariant under a global rescaling of the shear modulus of the shell ($\mu$-invariance) so that they can be computed in the fluid limit.
Since deviatoric Love numbers tend to zero in this limit, isostatic ratios are controlled by the partial derivatives of Love numbers with respect to the shear modulus of the shell (L'H\^opital's rule).
In particular, the shape ratio given by Eq.~(\ref{Sn}) (with a superscript  `e'  for `elastic') becomes
\begin{equation}
S_n^{\rm e} = \frac{\dot h_o^L + \zeta_n^\circ \, \dot h_o^I}{\dot h_s^L + \zeta_n^\circ \, \dot h_s^I} \, ,
\label{ShapeRatio0}
\end{equation}
where $\zeta_n^\circ$ denotes the fluid limit of the loading ratio and
\begin{equation}
\dot h^J_j = \frac{\partial}{\partial \mu_0} \, h^J_{j \rm e} \Big|_{\mu_0=0} \, .
\label{LovePartialE}
\label{LoadingRatioFluidLimit}
\end{equation}
If the shell has uniform elastic properties, $\mu_0$ is simply the shear modulus of the material; if elasticity varies with depth, $\mu_0$ is an arbitrary reference value in the factorization $\mu_s(r)=\mu_0f(r)$.
If the shell is stratified into $N$ homogeneous layers with shear moduli $(\mu_1,...,\mu_N)$, the partial derivative for the radial Love number can be expressed with the chain rule as
\begin{eqnarray}
\dot h^J_j &=& \sum_{i=1}^N \left. \left( \frac{\partial}{\partial \mu_i} \, h^J_{j \rm e} \right)  \left( \frac{\partial}{\partial \mu_0} \, \mu_i \right) \right|_{\mu_0=0}
\nonumber \\
&=& \sum_{i=1}^N  \left. f_i \, \frac{ \partial }{ \partial \mu_i } \, h^J_{j \rm e} \right|_{\mu_1=0 ... \mu_N=0} \, ,
\label{LovePartialDiscrete}
\end{eqnarray}
where $f_i$ is the discretized version of $f(r)$.
A similar expression holds for $\dot k^J_j$.
Eq.~(\ref{LovePartialDiscrete}) will be useful to prove the equivalence between zero deflection isostasy and `constant shape' viscoelastic isostasy.

The fluid loading ratios for zero deflection isostasy (ZDI) and for minimum stress isostasy (MSI) are given by
\begin{eqnarray}
\zeta_n^{ \circ \rm ZDI} &=& - \frac{\Delta\xi_1}{\xi_{s1}} \, \frac{x}{\alpha} \, ,
\label{zetaZDI0} \\
\zeta_n^{\circ \rm MSI} &=& \frac{1}{x} \, \frac{ \beta \left(1-a\right) + b}{ \beta \, c + d} \, ,
\label{zetaMING0}
\end{eqnarray}
where $(a,b,c,d)$ are the same as in Eqs.~(\ref{TnSnRelation})-(\ref{FnSnRelation}) (the MSI formula is derived under the assumption of an incompressible body with a homogenous shell).
The isostatic family parameters $\alpha$ and $\beta$ reflect the choice of boundary conditions.
For ZDI, $\alpha=0$ (resp.\ $\alpha=\infty$) if the radial displacement (or deflection) is zero at the top (resp.\ bottom) of the shell.
For MSI, $\beta=0$ (resp.\ $\beta=-\infty$) if the shape of the top (resp.\ bottom) of the shell is held constant when the energy is minimized.
MSI and ZDI isostatic ratios are equal if their loading ratios are equal, which occurs if
\begin{equation}
\alpha =  - \frac{b}{c} \, \frac{ \beta \, c + d }{  \beta \left(1-a\right) + b  } \, .
\label{alphaMSI}
\end{equation}
This duality can be used to obtain MSI isostatic ratios from ZDI ones, which are more easily computable.

If the body is incompressible and has a homogeneous shell, Love numbers can be expressed as ratio of degree-2 polynomials in $\mu_s$ with generic coefficients as in Eq.~(\ref{hgenV}).
In that case, the ZDI shape ratio reads
\begin{equation}
S_n^{\rm ZDI} = - \frac{\xi_{s1}}{x\Delta\xi_1} \, \frac{B_s^L + \alpha \, B_o^L}{B_s^I + \alpha \, B_o^I} \, .
\label{ShapeRatioZDIGen}
\end{equation}
The coefficients $B^J_j$ are given in analytical form in the complementary software for an incompressible body with homogeneous shell, ocean, and core.
The topographic ratio takes a particularly simple form which is independent of the properties of the core:
\begin{equation}
T_n^{\rm ZDI} = - \frac{\xi_{s1}}{\Delta\xi_1} \, \frac{1}{\gamma_o} \,  \frac{\alpha \, P_n(x) + Q_n(x)}{\alpha \, Q'_n(x) + x^2 \, P_n(x)} \, ,
\label{TopoRatioZDI}
\end{equation}
where the functions $(P_n(x),Q_n(x),Q'_n(x))$ are polynomials in $x$ with coefficients defined in Table~4 of Paper~I (see also complementary software).
In Paper~I, I  conjectured that the topographic ratio is in general independent of the internal structure below the shell, in which case Eq.~(\ref{TopoRatioZDI}) is valid for bodies with stratified core and stratified ocean as long as the shell is homogeneous and incompressible.
Using Eqs.~(\ref{TnSnRelation})-(\ref{FnSnRelation}), one can convert the topographic ratio into the shape ratio and compensation factor.

In the thin shell limit, the ZDI and MSI isostatic ratios do not depend on their respective isostatic family parameters $\alpha$ and $\beta$ up to first order in $\varepsilon=1-x$.
In particular, the shape ratio tends to
\begin{equation}
S_n^{\rm TSI} \cong - \frac{\xi_{s1}}{\Delta\xi_1} \left( 1 + \varepsilon \, \frac{n-1}{2n+1-3\xi_{o1}} \left( 3 \xi_{o1} - \frac{ \left(2n+1\right) \left(n+2\right)}{2n^2+2n-1} \right) \right) ,
\label{ShapeRatioTSI}
\end{equation}
where `TSI' stands for Thin Shell Isostasy (the core is assumed to be infinitely rigid).
Therefore the ZDI and MSI compensation factors do not depend on their isostatic family parameters to leading order in the thin shell limit.
While the thin shell limit is not a good approximation for the compensation factor except at the longest wavelengths, the compensation factor is very well approximated by computing it from the shape ratio (Eq.~(\ref{ShapeRatioTSI})) with Eq.~(\ref{FnSnRelation}).
This model of `improved' thin shell isostasy is very close to elastic isostasy with local compensation (see Fig.~10 of Paper~I).

%%%%%%%%%%%%%%%%%%%%%%%%%%%%%%%%%%%%%%%%%%%%%%%%%%%%%%%%%%%%%%%%%%%%%%%%%%%%%%%%%%%%%%%%%%%%%%%%%%%%%%%
%%%%%%%%%%%%%%%%%%%%%%%%%%%%%%%%%%%%%%%%%%%%%%%%%%%%%%%%%%%%%%%%%%%%%%%%%%%%%%%%%%%%%%%%%%%%%%%%%%%%%%%
%%%%%%%%%%%%%%%%%%%%%%%%%%%%%%%%%%%%%%%%%%%%%%%%%%%%%%%%%%%%%%%%%%%%%%%%%%%%%%%%%%%%%%%%%%%%%%%%%%%%%%%
\section{Constant loading of a homogeneous shell}
\label{AppendixPolesResidues}
\renewcommand{\theequation}{B.\arabic{equation}} % redefine the command that creates the equation no.
\setcounter{equation}{0}  % reset counter 

%%%%%%%%%%%%%%%%%%%%%%%%%%%%%%%%%%%%%%%%%%%%%%%%%%%%%%%%%%%%%%%%%%%%%%%%%%%%%%%%%%%%%%%%%%%%%%%%%%%%%%%
%%%%%%%%%%%%%%%%%%%%%%%%%%%%%%%%%%%%%%%%%%%%%%%%%%%%%%%%%%%%%%%%%%%%%%%%%%%%%%%%%%%%%%%%%%%%%%%%%%%%%%%
\subsection{Viscoelastic Love numbers}
\label{AppendixA1}

Consider an incompressible 3-layer body with a homogeneous shell (the ocean and the core are not necessarily homogeneous).
The shell is viscoelastic, the ocean is inviscid, and the core is elastic (including the limits in which the core is either infinitely rigid or fluid-like).
In that case, there are only two relaxation (buoyancy) modes associated with the shell boundaries (see Section 1.8 of \citet{sabadini2016}).
Two modes mean that the Love numbers have two poles in the Laplace domain.
Thus, Laplace-domain Love numbers can be written as the ratio of two degree-2 polynomials in the $s$-dependent shear modulus $\tilde \mu$ (the dependence on $s$ is implicit).
The principle of correspondence tells us the same thing: Laplace-domain Love numbers are given by elastic Love numbers (Eqs.~(D.1) and (D.12) of Paper~I) in which the elastic shear modulus is replaced by the $s$-dependent shear modulus $\tilde \mu$.
For example, viscoelastic radial Love numbers can be read from the generic form given by Eq.~(D.1) of Paper~I:
\begin{equation}
\tilde h_j^J(s) = \frac{A_j^J + B_j^J \tilde\mu(s) + C_j^J \tilde\mu(s)^2}{D + E \tilde\mu(s) + F \tilde\mu(s)^2} \, ,
\label{hgenV}
\end{equation}
The denominator coefficients are common to all Love numbers and determine the relaxation times.
The numerator coefficients $(A_j^J,B_j^J,C_j^J)$ satisfy Eqs.~(D.2)-(D.5) of Paper~I:
\begin{eqnarray}
A^J_j &=& D \, h_j^{J\circ} \, ,
\label{CoeffALS} \\
C^J_j &=& 0 \, ,
\label{CoeffCLS} \\
A_s^L \, B_o^I + A_o^I \, B_s^L &=& (E/D) \, A_s^L \, A_o^I \, ,
\label{IdentityAB} \\
B_s^L \, B_o^I - B_o^L \, B_s^I &=& (F/D) \, A_s^L \, A_o^I \, .
\label{IdentityBB} 
\end{eqnarray}
The first two constraints come from the limits in which the shell tends to be fluid-like or infinitely rigid.
The last two come from the assumption of $\mu$-invariance of zero-deflection elastic isostasy (they were checked analytically for a model with three homogeneous layers).

%%%%%%%%%%%%%%%%%%%%%%%%%%%%%%%%%%%%%%%%%%%%%%%%%%%%%%%%%%%%%%%%%%%%%%%%%%%%%%%%%%%%%%%%%%%%%%%%%%%%%%%
%%%%%%%%%%%%%%%%%%%%%%%%%%%%%%%%%%%%%%%%%%%%%%%%%%%%%%%%%%%%%%%%%%%%%%%%%%%%%%%%%%%%%%%%%%%%%%%%%%%%%%%
\subsection{Poles and decay times}

The two (negative) $\tilde\mu$-poles of Eq.~(\ref{hgenV}) are given by
\begin{equation}
\mu_\pm = \frac{1}{2F} \left( - E \pm\sqrt{E^2-4\,D F} \, \right) \, ,
\label{mupm}
\end{equation}
and satisfy the identities
\begin{equation}
\mu_+\mu_-=D/F
\hspace{5mm} \mbox{and} \hspace{5mm}
\mu_++\mu_-=-E/F \, .
\label{PoleID}
\end{equation}
For Maxwell rheology (Eq.~(\ref{maxwell})), the corresponding (negative) $s$-poles are given by
\begin{equation}
\mu_\pm = \frac{\mu_{\rm e} \, s_\pm}{s_\pm+\mu_{\rm e}/\eta}
\,\, \Leftrightarrow \,\,
s_\pm = - \frac{\mu_{\rm e}}{\eta} \, \frac{\mu_\pm}{\mu_\pm-\mu_{\rm e}} \, .
\label{spoles}
\end{equation}
The viscoelastic decay times are given by
\begin{equation}
\tau_\pm= -1/s_\pm \, ,
\label{DecayTimesViscoelastic}
\end{equation}
with the longest time scale corresponding to $s_+$, the pole closest to zero.
Viscoelastic decay times reduce to viscous decay times (Eq.~(\ref{ViscousDecayTimes})) in the limit of infinite elastic modulus:
\begin{equation}
\tau_\pm^{\rm viscous}= -\eta/\mu_\pm \, .
\label{DecayTimesEquiv}
\end{equation}
This property can be proven by showing that the viscous decay times are the solutions of the following equation (obtained from  $D+E\mu_\pm+F\mu_\pm^2=0$):
\begin{equation}
D \, (-\tau_\pm^{\rm viscous}/\eta)^2 + E \, (-\tau_\pm^{\rm viscous}/\eta) + F = 0 \, .
\end{equation}

%%%%%%%%%%%%%%%%%%%%%%%%%%%%%%%%%%%%%%%%%%%%%%%%%%%%%%%%%%%%%%%%%%%%%%%%%%%%%%%%%%%%%%%%%%%%%%%%%%%%%%%
%%%%%%%%%%%%%%%%%%%%%%%%%%%%%%%%%%%%%%%%%%%%%%%%%%%%%%%%%%%%%%%%%%%%%%%%%%%%%%%%%%%%%%%%%%%%%%%%%%%%%%%
\subsection{Residues}

The residues associated with the poles $s_\pm$ are given by
\begin{equation}
r^J_{j\pm} = \Big( \frac{\mu_{\rm e} \, s_\pm}{\mu_\pm} \Big)^2 \, \frac{1}{s_\pm - s_\mp} \, \frac{A_j^J + B_j^J \mu_\pm + C_j^J \mu_\pm^2}{D + E \mu_{\rm e} + F \mu_{\rm e}^2} \, .
\end{equation}
All factors except the numerator of the last term are the same whatever the Love number.
These common factors cancel when computing the ratio of residues as required for the shape ratio and compensation factor (Eqs.~(\ref{SnVeLI})-(\ref{FnVeLI})).
The shape ratio can be computed either with top loading or from bottom loading:
\begin{eqnarray}
S_n^{{\rm VeLI, top}} &=& \frac{ r^L_{o+}}{r^L_{s+}} \,\, = \,\, \frac{B_o^L \, \mu_+}{A_s^L + B_s^L \, \mu_+} \, ,
\label{SnVeLItop} \\
S_n^{\rm VeLI, bot} &=&  \frac{r^I_{o+}}{r^I_{s+}} \,\, = \,\, \frac{A_o^I + B_o^I \, \mu_+}{B_s^I \, \mu_+} \, .
\label{SnVeLIbot}
\end{eqnarray}
These ratios do not depend on $\mu_{\rm e}$ or $\eta$.
Moreover, the two ratios are identical if
\begin{equation}
A_s^L \, A_o^I + \left( A_s^L \, B_o^I + A_o^I \, B_s^L \right) \mu_+ + \left( B_s^L \, B_o^I - B_o^L \, B_s^I \right) \mu_+^2 = 0 \, ,
\label{IdentityConstantLoad}
\end{equation}
which is true because this equation is proportional to $D+E\mu_++F\mu_+^2=0$ thanks to the identities (\ref{IdentityAB})-(\ref{IdentityBB}).
Therefore, $\mu$-invariance implies that viscoelastic isostasy with constant load does not depend of the loading ratio.

For further analysis, the shape ratio is set into equivalent expressions where $\mu_+$ appears either only in the denominator or only in the numerator:
\begin{eqnarray}
S_n^{\rm VeLI} &=& \frac{\xi_{sn} \, B_o^L }{F \mu_+ - x\Delta\xi_n \, B_o^I}
\label{SnVeLIAnalyticalTop} \\
&=& \frac{F \mu_+ - \xi_{sn} \, B_s^L}{x\Delta\xi_n \, B_s^I } \, .
\label{SnVeLIAnalyticalBot}
\end{eqnarray}
These formulas were obtained by multiplying Eqs.~(\ref{SnVeLItop})-(\ref{SnVeLIbot}) by $\mu_-/\mu_-$ and using Eqs.~(\ref{CoeffALS}), (\ref{IdentityAB}), and (\ref{PoleID}).

%%%%%%%%%%%%%%%%%%%%%%%%%%%%%%%%%%%%%%%%%%%%%%%%%%%%%%%%%%%%%%%%%%%%%%%%%%%%%%%%%%%%%%%%%%%%%%%%%%%%%%%
%%%%%%%%%%%%%%%%%%%%%%%%%%%%%%%%%%%%%%%%%%%%%%%%%%%%%%%%%%%%%%%%%%%%%%%%%%%%%%%%%%%%%%%%%%%%%%%%%%%%%%%
\subsection{Thin shell limit}

Recall that the thin shell expansion parameter is $\varepsilon=1-x=d/R$.
In Paper~I, I showed that the coefficients $D$ and $F$ decrease faster with shell thickness than $E$, whereas $B^J_j$ decreases as fast as $E$.
Given that $B^J_j/E\rightarrow-1/\xi_{on}$ in the thin shell limit (Eq.~(D.21) of Paper~I), the four coefficients $B_j^J$ start to differ at ${\cal O}(\varepsilon^2)$.
Suppose that $E$ is ${\cal O}(\varepsilon)$, as is the case in my explicit computation (all coefficients can of course be multiplied by the same power of $1-x=\varepsilon$); then $B^J_j$ is ${\cal O}(\varepsilon)$ whereas $D$ and $F$ are ${\cal O}(\varepsilon^2)$.
Thus, the $\mu$-poles (Eq.~(\ref{mupm})) can be approximated to leading order in $\varepsilon$ by
\begin{equation}
\left( \mu_+ , \mu_- \right) \cong - \left( \frac{D}{E} , \frac{E}{F} \right) .
\label{mupmA}
\end{equation}
If the ocean is homogeneous and the core is infinitely rigid, the expansions of $D/E$ and $E/F$ to order $\varepsilon$ are given by (in nondimensional form)
\begin{eqnarray}
\bar\mu_+ &\cong& - \frac{ n \left(n+1\right)}{2\left(2n^2+2n-1\right)} \, \frac{\xi_{s1}}{\xi_{o1}} \, \Delta\xi_1 \, \varepsilon \, + \, {\cal O}(\varepsilon^2) \, ,
\label{muplusTS} \\
\bar\mu_- &\cong& - \frac{2n^2+2n-1}{6\left(n-1\right)\left(n+2\right)} \, \xi_{o1} \left(1-\xi_{on} \right) \frac{1}{\varepsilon} \, + \, {\cal O}(1) \, ,
\label{muminTS}
\end{eqnarray}
where $\bar \mu_\pm=\mu_\pm/\mu_G$, $\mu_G$ being the gravitational rigidity $\mu_G=\rho_sg_sR_s$.
Eq.~(\ref{muplusTS}), but not Eq.~(\ref{muminTS}), remains correct if the core is homogeneous and elastic.

Thus $|\bar\mu_+|\ll1$ if the shell is thin.
For large satellites, $\mu_{\rm e}$ has a magnitude comparable to $\mu_G$, whereas $\mu_{\rm e}\gg\mu_G$ for small satellites.
Therefore $|\mu_+|\ll\mu_{\rm e}$ for physically plausible values of $\mu_{\rm e}$.
Because of this, the $s$-pole closest to zero (Eq.~(\ref{spoles})) can be approximated by
\begin{equation}
s_+ \cong \frac{\mu_+}{\eta} \, .
\label{splusapprox}
\end{equation}
This value is independent of $\mu_{\rm e}$ and is inversely proportional to the longest viscous decay time (Eq.~(\ref{DecayTimesEquiv})).

The thin shell limit of the shape ratio is most conveniently obtained from Eqs.~(\ref{SnVeLIAnalyticalTop})-(\ref{SnVeLIAnalyticalBot}) given that $F\mu_+\sim{\cal O}(\varepsilon^3)$ can be dropped from the equation:
\begin{eqnarray}
S_n^{\rm VeLI} &\cong& - \frac{\xi_{s1}}{x\Delta\xi_1} \, \frac{B_o^L }{B_o^I} + {\cal O}(\varepsilon^2)
\label{SnVeLIthinshell1} \\
&\cong& - \frac{\xi_{s1}}{x\Delta\xi_1} \, \frac{B_s^L }{B_s^I} + {\cal O}(\varepsilon^2) \, .
\label{SnVeLIthinshell2}
\end{eqnarray}
These expressions coincide with the ZDI shape ratio in which either $\alpha=0$ or $\alpha=\infty$ (Eq.~(\ref{ShapeRatioZDIGen})).
Thus, the expansion to ${\cal O}(\varepsilon)$ of the viscoelastic `constant load' shape ratio must agree with the expansion to ${\cal O}(\varepsilon)$ of the elastic shape ratio.
The explicit thin shell limit of the shape ratio is given by Eq.~(\ref{ShapeRatioTSI}) under the assumption of an infinitely rigid core.

%%%%%%%%%%%%%%%%%%%%%%%%%%%%%%%%%%%%%%%%%%%%%%%%%%%%%%%%%%%%%%%%%%%%%%%%%%%%%%%%%%%%%%%%%%%%%%%%%%%%%%%
%%%%%%%%%%%%%%%%%%%%%%%%%%%%%%%%%%%%%%%%%%%%%%%%%%%%%%%%%%%%%%%%%%%%%%%%%%%%%%%%%%%%%%%%%%%%%%%%%%%%%%%
\subsection{Shape ratio at high harmonic degree}
\label{AppendixAsymptotic}

At high harmonic degree, the viscoelastic shape ratio for a homogeneous shell depends on the harmonic degree $n$, on the relative radius of shell-ocean boundary $x$, and on the parameter $\psi$ defined by
\begin{equation}
\psi = \gamma_o \, x \, \frac{\Delta\xi_1}{\xi_{s1}} \, ,
\label{defpsi}
\end{equation}
with $\psi\leq1$ being the physically most likely case.
The asymptotic behaviour of the `constant load' viscoelastic shape ratio at high harmonic degree is given by
\begin{eqnarray}
S_n^{\rm VeLI} &\sim& - \frac{1}{n \, x^n} \, \frac{1}{1-x^2} \left( \frac{1}{\psi} -1 \right)
\hspace{5mm} \mbox{if $\psi\leq1$} \, , 
\label{SnVeLIAsymptoticTop} \\
S_n^{\rm VeLI} &\sim& -n \, x^n \, \frac{1-x^2}{x} \, \frac{1}{\psi-1}
\hspace{12mm} \mbox{if $\psi>1$}  \, .
\label{SnVeLIAsymptoticBot} 
\end{eqnarray}
The formulas for $\psi\leq1$ and $\psi>1$ were obtained by substituting the expressions of Table~\ref{TableAsymptotic} into Eq.~(\ref{SnVeLIbot}) and Eq.~(\ref{SnVeLItop}), respectively (doing the other way around is tricky because dominant terms cancel at the leading asymptotic order).

The asymptotic behaviour of the `constant shape' viscoelastic shape ratio at high harmonic degree is given by
\begin{eqnarray}
S_n^{\rm VeSI, top} &\sim& - n \, x^{n} \, \frac{1-x^2}{x} \, \frac{1}{\psi} \, ,
\label{SnVeSIAsymptoticTop} \\
S_n^{\rm VeSI, bot} &\sim& - \frac{1}{n \, x^n} \, \frac{1}{1-x^2} \, \frac{1}{\psi} \, .
\label{SnVeSIAsymptoticBot}
\end{eqnarray}
These formulas were obtained by substituting the expressions of Table~\ref{TableAsymptotic} into Eq.~(\ref{ShapeRatioVeSIGen}) with either $\zeta^{\rm V}_n=0$ (top load) or $\zeta^{\rm V}_n=\pm\infty$ (bottom load).

Thus, $S_n^{\rm VeLI}$ tends either to a constant multiple of $S_n^{\rm VeSI, bot}$ or of $S_n^{\rm VeSI, top}$, depending on the value of $\psi$:
\begin{eqnarray}
S_n^{\rm VeLI} &\sim& \left( 1- \psi \right) S_n^{\rm VeSI, bot}
\hspace{5mm} \mbox{if $\psi\leq1$} \, ,
\label{SnVeLIAsymptoticRatio1} \\
S_n^{\rm VeLI} &\sim& \frac{\psi}{\psi-1} \,\, S_n^{\rm VeSI, top}
\hspace{7mm} \mbox{if $\psi>1$}  \, .
\label{SnVeLIAsymptoticRatio2}
\end{eqnarray}

%TABLE
\begin{table}[h]\centering
\ra{1.3}
\small
\caption[High-degree asymptotic behaviour of Love number generic coefficients]{High-degree asymptotic behaviour of Love number generic coefficients and $\mu$-pole closest to zero
for a 3-layer incompressible body with homogeneous layers (the core can be rigid, fluid, or elastic).
The parameter $\psi$ is defined by Eq.~(\ref{defpsi}) while fluid Love numbers $h^{J\circ}_j$ are given in Table~\ref{TableLoveV}.
}
\vspace{1.5mm}
\begin{tabular}{@{}ll@{}}
\toprule
Coeff. & \hspace{1.5mm} Asymptotic behaviour \\
%\hline
\midrule
$D/F$ & \hspace{1.5mm} $(\xi_{s1})^2 \, \psi/(4n^2)$ \\
$E/F$ &  \hspace{1.5mm} $\xi_{s1} \, (1 + \psi)/(2n)$ \\
$A^J_j/F$ & \hspace{1.5mm}  $h^{J\circ}_j \, D/F$ \\
$B^L_s/F$ & $-1/3$ \\
$B^L_o/F$ & $-n \, x^{n-1} \, (1-x^2)/3$ \\
$B^I_s/F$ & $-n \, x^n (1-x^2) \, \gamma_o/3$ \\
$B^I_o/F$ & $-\gamma_o/3$ \\
$\mu_+$ & $-\xi_{s1}(1+\psi-|1-\psi|)/(4n)$ \\
$\mu_+ (\psi\leq1)$ & $-\xi_{s1} \, \psi/(2n)$ \\
$\mu_+ (\psi\geq1)$ & $-\xi_{s1}/(2n)$ \\
%\hline
\addlinespace[2pt]
\bottomrule
\end{tabular}
\label{TableAsymptotic}
\end{table}%

%%%%%%%%%%%%%%%%%%%%%%%%%%%%%%%%%%%%%%%%%%%%%%%%%%%%%%%%%%%%%%%%%%%%%%%%%%%%%%%%%%%%%%%%%%%%%%%%%%%%%%%
%%%%%%%%%%%%%%%%%%%%%%%%%%%%%%%%%%%%%%%%%%%%%%%%%%%%%%%%%%%%%%%%%%%%%%%%%%%%%%%%%%%%%%%%%%%%%%%%%%%%%%%
%%%%%%%%%%%%%%%%%%%%%%%%%%%%%%%%%%%%%%%%%%%%%%%%%%%%%%%%%%%%%%%%%%%%%%%%%%%%%%%%%%%%%%%%%%%%%%%%%%%%%%%
\section{Slow viscous flow}
\label{AppendixViscousFlow}
\renewcommand{\theequation}{C.\arabic{equation}} % redefine the command that creates the equation no.
\setcounter{equation}{0}  % reset counter

%%%%%%%%%%%%%%%%%%%%%%%%%%%%%%%%%%%%%%%%%%%%%%%%%%%%%%%%%%%%%%%%%%%%%%%%%%%%%%%%%%%%%%%%%%%%%%%%%%%%%%%
%%%%%%%%%%%%%%%%%%%%%%%%%%%%%%%%%%%%%%%%%%%%%%%%%%%%%%%%%%%%%%%%%%%%%%%%%%%%%%%%%%%%%%%%%%%%%%%%%%%%%%%
\subsection{Differential equations}

Slow viscous flow is described by 6 variables with the dimension of velocity, which can be grouped into two sets \citep{hager1989}.
The first set, mainly characterized by velocity and stress, reads $\mathbf{u}=(u_1, u_2, u_3, u_4)^T$, where $u_1$ and $u_2$ are the radial and tangential velocities, $u_3=ry_3/\eta_0+(\rho/\eta_0)ry_5$ is related to the (Eulerian) radial stress $y_3$ and the gravitational potential $y_5$, and $u_4=ry_4/\eta_0$ is the potential for the (Eulerian) radial-tangential stress.
The second set, including only gravity variables, reads $\mathbf{v}=(v_1 , v_2)^T$ where $v_1=(\rho_0/\eta_0)ry_5$ and $v_2=(\rho_0/\eta_0)r^2y_6$ are associated with the gravity potential and its derivative, respectively.
A reference density $\rho_0$ and a reference viscosity $\eta_0$ have been introduced for normalization purposes, but the solution should not depend on them.
In a homogeneous layer, the 6 variables satisfy
\begin{eqnarray}
r \, \frac{d \mathbf{u} }{dr} &=& \mathbf{A \cdot u} \, ,
\label{Asystem} \\
r \, \frac{d \mathbf{v} }{dr}&=& \mathbf{B \cdot v} \, ,
\label{Bsystem}
\end{eqnarray}
where
\begin{eqnarray}
\mathbf{A} &=& \left(
\begin{array}{cccc}
-2 & -\delta_n & 0 & 0  \\
-1 & 1 & 0 & 1/\eta^* \\
12\eta^* & 6\eta^*\delta_n & 1 & -\delta_n \\
-6\eta^* & -2\eta^*(2\delta_n+1) & -1 & -2 \\
\end{array}
\right) ,
\label {ViscousMatA} \\
\mathbf{B} &=& \left(
\begin{array}{cc}
1 & 1  \\
-\delta_n & 0 \\
\end{array}
\right) ,
\label {ViscousMatB}
\end{eqnarray}
in which $\eta^*=\eta/\eta_0$ and $\delta_n=-n(n+1)$ as elsewhere.

%%%%%%%%%%%%%%%%%%%%%%%%%%%%%%%%%%%%%%%%%%%%%%%%%%%%%%%%%%%%%%%%%%%%%%%%%%%%%%%%%%%%%%%%%%%%%%%%%%%%%%%
%%%%%%%%%%%%%%%%%%%%%%%%%%%%%%%%%%%%%%%%%%%%%%%%%%%%%%%%%%%%%%%%%%%%%%%%%%%%%%%%%%%%%%%%%%%%%%%%%%%%%%%
\subsection{Fundamental matrices}

The systems of differential equations (\ref{Asystem}) and (\ref{Bsystem}) can be solved with the \textit{(Eulerian) propagator matrix method}, in which the propagator matrix relates the variables at two different radii within the layer.
\citet{hager1989} compute the propagator matrices with Sylvester's formula, but without giving the result for $\mathbf{A}$ (because it is too long), while there are typos in their propagator matrix for $\mathbf{B}$. 
Instead of computing directly the propagator matrix, I will first solve the equations in terms of free constants.
In this way, I can express more compactly the propagator in terms of the \textit{fundamental matrix}, as is commonly done in gravitational-elastic theory.
If $\mathbf{u}\sim r^p$ and $\mathbf{v}\sim r^q$, then Eqs.~(\ref{Asystem})-(\ref{Bsystem}) are transformed into eigenvalue problems:
\begin{eqnarray}
p \, \mathbf{u} &=& \mathbf{A \cdot u} \, ,
\\
q \, \mathbf{v} &=& \mathbf{B \cdot v} \, .
\label{eigensystems}
\end{eqnarray}
The eigenvalues of $\mathbf{A}$ and $\mathbf{B}$ are $(n+1,n-1,-n,-(n+2))$ and $(n+1,-n)$, respectively.
The solutions can be written as a linear combination of the eigenvectors multiplied by the corresponding  $r^p$ or $r^q$ or, in matrix form:
\begin{eqnarray}
\mathbf{u}(r) &=& \mathbf{Y}_A(r) \cdot \mathbf{c}_u \, ,
\label{uYA}\\
\mathbf{v}(r) &=& \mathbf{Y}_B(r) \cdot \mathbf{c}_v \, ,
\label{vYB}
\end{eqnarray}
where $\mathbf{c}_u$ and $\mathbf{c}_v$ are vectors of 4 free constants and 2 free constants, respectively.
The fundamental matrices are given by
\begin{eqnarray}
\mathbf{Y}_A(r) &=&  \mathbf{C}_A \cdot \mathbf{\bar Y}_A \cdot \mathbf{D}_A(r) \, ,
\label{YA} \\
\mathbf{Y}_B(r) &=& \mathbf{\bar Y}_B \cdot \mathbf{D}_B(r) \, ,
\label{YB}
\end{eqnarray}
where
\begin{eqnarray}
\mathbf{\bar Y}_A &=& \left(
\begin{array}{cccc}
1 & 1 & 1 & 1 \\
\frac{n+3}{n(n+1)} & \frac{1}{n} & -\frac{n-2}{n(n+1)} & -\frac{1}{n+1} \\
\frac{n^2-n-3}{n} & n-1 & -\frac{n^2+3n-1}{n+1} & -(n+2)\\
\frac{n+2}{n+1} & \frac{n-1}{n} & \frac{n-1}{n} & \frac{n+2}{n+1} \\
\end{array}
\right) ,
\label{YAbar} \\
\mathbf{\bar Y}_B &=& \left(
\begin{array}{cccc}
1 & 1  \\
n & -(n+1) \\
\end{array}
\right) ,
\label{YBbar}
\end{eqnarray}
while $ \mathbf{C}_A$, $\mathbf{D}_A(r)$, and $\mathbf{D}_B(r)$ are diagonal matrices with elements
\begin{eqnarray}
{\rm diag}( \mathbf{C}_A ) &=& \left(1 , 1 , 2\eta^* ,  2\eta^* \right) ,
\\
{\rm diag}( \mathbf{D}_A(r) ) &=& \left({\bar r}^{n+1} , {\bar r}^{n-1} , {\bar r}^{-n} ,  {\bar r}^{-n-2} \right) ,
\label{DA} \\
{\rm diag}(\mathbf{D}_B(r) ) &=& \left( {\bar r}^{n+1} , {\bar r}^{-n} \right) ,
\label{DB}
\end{eqnarray}
where $\bar r=r/r_0$ ($r_0$ is an arbitrary reference radius introduced so that the fundamental matrix is nondimensional).
The solution for $\mathbf{u}$ is similar to Eq.~(22) of \citet{ricard1984}, but without the typos in the 3rd term of $u_2$ and in the 3rd and 4rth terms of $u_3$ (note that the solution without self-gravity of \citet{ricard1984} is formally equivalent to the solution $\mathbf{u}$ of \citet{hager1989} because of gravity decoupling in the latter paper).

The inverse fundamental matrices read
\begin{eqnarray}
\mathbf{Y}_A^{-1}(r) &=&  \mathbf{D}_A^{-1}(r) \cdot \mathbf{\bar Y}_A^{-1} \cdot \mathbf{C}_A^{-1} \, ,
\label{YAInv} \\
\mathbf{Y}_B^{-1}(r) &=& \mathbf{D}_B^{-1}(r) \cdot \mathbf{\bar Y}_B^{-1} \, ,
\label{YBInv}
\end{eqnarray}
where
\begin{eqnarray}
\mathbf{\bar Y}_A^{-1} &=&
\mathbf{E}_A \cdot
\left(
\begin{array}{cccc}
-(n+2) & n(n+2) & -1 & n \\
\frac{n^2+3n-1}{n+1} & - (n^2-1) & 1 & -(n-2) \\
n-1 & n^2-1 & -1 & -(n+1) \\
-\frac{n^2-n-3}{n} & -n(n+2) & 1 & n+3 \\
\end{array}
\right) ,
\label{YAbarInv} \\
\mathbf{\bar Y}_B^{-1} &=&
\frac{1}{2n+1}
\left(
\begin{array}{cccc}
n+1 & 1  \\
n & -1 \\
\end{array}
\right) ,
\label{YBbarInv}
\end{eqnarray}
in which $\mathbf{E}_A$ is a diagonal matrix with elements
\begin{equation}
{\rm diag}( \mathbf{E}_A ) =  \frac{n(n+1)}{2n+1} \left( \frac{1}{2n+3} , \frac{1}{2n-1} , \frac{1}{2n-1} ,  \frac{1}{2n+3} \right) .
\end{equation}

%%%%%%%%%%%%%%%%%%%%%%%%%%%%%%%%%%%%%%%%%%%%%%%%%%%%%%%%%%%%%%%%%%%%%%%%%%%%%%%%%%%%%%%%%%%%%%%%%%%%%%%
%%%%%%%%%%%%%%%%%%%%%%%%%%%%%%%%%%%%%%%%%%%%%%%%%%%%%%%%%%%%%%%%%%%%%%%%%%%%%%%%%%%%%%%%%%%%%%%%%%%%%%%
\subsection{Propagator matrices}

Using Eqs.~(\ref{uYA})-(\ref{vYB}) twice and eliminating the free constants, I can relate the variables at radius $R_s$ to those at radius $R_o$,
\begin{eqnarray}
\mathbf{u}(R_s) &=& \mathbf{P}_A(R_s/R_o) \cdot \mathbf{u}(R_o) \, ,
\label{propagAappendix} \\
\mathbf{v}(R_s) &=& \mathbf{P}_B(R_s/R_o) \cdot \mathbf{v}(R_o) \, ,
\label{propagBappendix}
\end{eqnarray}
where $\mathbf{P}_A$ and $\mathbf{P}_A$ are the propagator matrices:
\begin{eqnarray}
\mathbf{P}_A(R_s/R_o) &=& \mathbf{Y}_A(R_s) \cdot \mathbf{Y}_A^{-1}(R_o) \, ,
\label{PA} \\
\mathbf{P}_B(R_s/R_o) &=&  \mathbf{Y}_B(R_s) \cdot \mathbf{Y}_B^{-1}(R_o) \, .
\label{PB}
\end{eqnarray}
For example, the  propagator matrix for $\mathbf{B}$ is given by
\begin{equation}
\mathbf{P}_B(R_s/R_o) = \frac{1}{2n+1}
\left(
\begin{array}{ll}
\left(n+1\right) x^{-n-1} + n \, x^n &x^{-n-1} - x^n  \\
n\left(n+1\right) \left(x^{-n-1}-x^n\right) & n \, x^{-n-1} + \left(n+1\right) x^n
\end{array}
\right) ,
\label{PBexplicit}
\end{equation}
where $x=R_o/R_s$.
This formula corrects typos in Eq.~(4.44) of \citet{hager1989}.
It is straightforward to compute $\mathbf{P}_A$ as the two-point product of the fundamental matrix and its inverse, but the result is quite long and will not be given here.

%%%%%%%%%%%%%%%%%%%%%%%%%%%%%%%%%%%%%%%%%%%%%%%%%%%%%%%%%%%%%%%%%%%%%%%%%%%%%%%%%%%%%%%%%%%%%%%%%%%%%%%
%%%%%%%%%%%%%%%%%%%%%%%%%%%%%%%%%%%%%%%%%%%%%%%%%%%%%%%%%%%%%%%%%%%%%%%%%%%%%%%%%%%%%%%%%%%%%%%%%%%%%%%
\subsection{Boundary conditions}

The boundary conditions for the radial-tangential stress variable $u_4$ are very simple (free-slip):
\begin{equation}
u_4(R_o) = u_4(R_s) = 0 \, .
\label{BCu4}
\end{equation}
The variable $u_3$ is a combination of the radial stress and the gravity potential.
The bottom boundary condition for $u_3$ does not appear explicitly in \citet{hager1989} (see their Eq.~(4.53)), while the surface boundary condition contains typos (see after their Eq.~(4.54)).
Here are the appropriate boundary conditions:
\begin{eqnarray}
u_3(R_s) &=& - \rho_s \left( g_s R_s \, \delta{}r_s/\eta_0 - v_1(R_s)/\rho_0 \right) \, ,
\label{BCu3s} \\
u_3(R_o) &=& \left(\rho_o-\rho_s\right) \left( g_o R_o \, \delta{}r_o/\eta_0 -  v_1(R_o)/\rho_0 \right) ,
\label{BCu3o}
\end{eqnarray}
where $\delta{}r_j$ is the displacement of interface $R_j$.
The boundary conditions for the derivative of the gravity potential at harmonic degree $n$ are given by Eqs.~(4.56) and (4.58) of \citet{hager1989}:
\begin{eqnarray}
v_2(R_s) &=& -\left(n+1\right) v_1(R_o) + 4\pi G R_s^2 \, \rho_s (\rho_0/\eta_0) \, \delta{}r_s \, ,
\label{BCv2s} \\
v_2(R_o) &=& n \, v_1(R_o) - 4\pi G R_o^2 \left(\rho_o-\rho_s\right) \left(\rho_0/\eta_0\right) \delta{}r_o \, .
\label{BCv2o}
\end{eqnarray}
Solving Eq.~(\ref{propagBappendix}) for $v_1$ with the appropriate boundary conditions for $v_2$ (Eqs.~(\ref{BCv2s})-(\ref{BCv2o})) and using nondimensional parameters, I get boundary conditions for the gravity potential variable in terms of boundary displacements:
\begin{eqnarray}
v_1(R_s) &=& \xi_{sn} \left(\delta{}r_s/\tau_0\right)  + \Delta\xi_n \, x^{n+2}  \left(\delta{}r_o/\tau_0\right) ,
\label{BCv1s} \\
v_1(R_o) &=&  \xi_{sn} \, x^{n+1} \left( \delta{}r_s/\tau_0 \right) + \Delta\xi_n \, x^2  \left(\delta{}r_o/\tau_0 \right) ,
\label{BCv1o}
\end{eqnarray}
where $\tau_0=\eta_0/(\rho_0g_sR_s)$ is a reference time.
These results are equivalent to Eqs.~(C.1)-(C.2) of Paper~I, giving the gravitational perturbations in terms of the shapes of the shell boundaries (noting that $H_{cn}=0$ because the core is rigid, and that the definition of $v_1$ introduces an additional factor of $x$).

Substituting Eqs.~(\ref{BCv1s})-(\ref{BCv1o}) into Eqs.~(\ref{BCu3s})-(\ref{BCu3o}) and using nondimensional parameters, I get boundary conditions for stress variables in terms of boundary displacements:
\begin{eqnarray}
u_3(R_s) &=& \xi_{s1} \left( \left( \xi_{sn} -1 \right) \left(\delta{}r_s/\tau_0\right) + \Delta\xi_n \, x^{n+2} \left( \delta{}r_o/\tau_0\right) \right) \, ,
\label{BCu3sBD} \\
u_3(R_o) &=& \Delta\xi_1 \left( - \xi_{sn} \, x^{n+1} \left( \delta{}r_s/\tau_0 \right) + x \left( \gamma_o - \Delta\xi_n \, x \right) \left( \delta{}r_o/\tau_0\right) \right) \, ,
\label{BCu3oBD}
\end{eqnarray}
where the reference density $\rho_0$ is set equal to the bulk density $\rho_b$ so that
\begin{equation}
\tau_0=\eta_0/(\rho_bg_sR_s)\, .
\label{deftau0}
\end{equation}

%%%%%%%%%%%%%%%%%%%%%%%%%%%%%%%%%%%%%%%%%%%%%%%%%%%%%%%%%%%%%%%%%%%%%%%%%%%%%%%%%%%%%%%%%%%%%%%%%%%%%%%
%%%%%%%%%%%%%%%%%%%%%%%%%%%%%%%%%%%%%%%%%%%%%%%%%%%%%%%%%%%%%%%%%%%%%%%%%%%%%%%%%%%%%%%%%%%%%%%%%%%%%%%
\subsection{Reduced propagation system}
\label{ReducedPropagationSystem}

Following \citet{ermakov2017}, I start by solving the propagation equation for the gravity variables for a shell of uniform density,
\begin{equation}
\left(
\begin{array}{c}
v_1(R_s) \\
v_2(R_s)
\end{array}
\right)
= \mathbf{P}_B \cdot
\left(
\begin{array}{c}
v_1(R_o) \\
v_2(R_o)
\end{array}
\right) ,
\label{propagB}
\end{equation}
where $\mathbf{P}_B$ is the matrix given by Eq.~(\ref{PBexplicit}).
In this way, the gravity potential can be expressed in terms of boundary displacements,
making it possible to express the boundary conditions on the radial stress variable $u_3$ in terms of boundary displacements (Eqs.~(\ref{BCv1s})-(\ref{BCu3oBD})).
The boundary conditions on the tangential-radial stress variable $u_4$ are free-slip (Eq.~(\ref{BCu4})).
The propagation of the velocity variables $(u_1,u_2)$ and stress variables $(u_3,u_4)$ from the bottom of the shell to the surface reads 
\begin{eqnarray}
\left(
\begin{array}{c}
u_1(R_s) \\
... \\
u_4(R_s)
\end{array}
\right)
= \mathbf{P}_A \cdot
\left(
\begin{array}{c}
u_1(R_o) \\
... \\
u_4(R_o)
\end{array}
\right) ,
\label{propagA}
\end{eqnarray}
where $\mathbf{P}_A$ is the two-point matrix product defined by Eq.~(\ref{PA}).
For our isostatic problem, this $4\times4$ matrix system can be reduced to two equations:
\begin{enumerate}
\item Drop the equation for the surface tangential velocity $u_2(R_s)$ from the system because isostatic ratios depend only on radial displacements and velocities.
\item Substitute the boundary conditions for $u_3$ and $u_4$ into the propagation system.
\item Eliminate the bottom tangential velocity $u_2(R_o)$ from the propagation system by using the equation for $u_4(R_s)$.
\end{enumerate}
This procedure yields two equations (or reduced propagation system) for the radial velocities $(u_1(R_s),u_1(R_o))$ and the radial displacements $(\delta{}r_s,\delta{}r_o)$.

\end{appendices}

\newpage

\bibliographystyle{agufull04}
\renewcommand{\baselinestretch}{0.5}

\end{document}